\documentclass[longauth]{aa}

\usepackage{graphicx}
\usepackage{natbib}         
\usepackage{scalerel}
\usepackage{lastpage}
\usepackage[table]{xcolor}
\newcommand{\diff}{\mathop{}\!\mathrm{d}}

\usepackage{booktabs}
\usepackage{tabularx}
\usepackage{siunitx}
\usepackage{multirow}
\usepackage{siunitx}

\usepackage{txfonts}
\usepackage[pdfencoding=auto,psdextra]{hyperref}
\hypersetup{
    colorlinks=true,
    linkcolor=blue,
    filecolor=magenta,      
    urlcolor=blue,
    citecolor=blue
}
\usepackage[nameinlink,capitalise]{cleveref}
\crefname{section}{Sect.}{Sects.}
\Crefname{section}{Section}{Sections}
\crefname{figure}{Fig.}{Figs.}
\Crefname{figure}{Figure}{Figures}
\crefname{equation}{Eq.}{Eqs.}
\Crefname{equation}{Equation}{Equations}
\crefname{table}{Table}{Tables}
\crefname{appendix}{Appendix}{Appendices}

\makeatletter
\def\aa@extracttenauthor#1\and#2\and#3\and#4\and#5\and#6\and#7\and#8\and#9\aa@nil{%
  \aa@splitninth#9\and\aa@nil
  \gdef\aa@tenauthors{%
    #1\unskip, #2\unskip, #3\unskip, #4\unskip, #5\unskip, #6\unskip, #7\unskip, #8\unskip, \aa@ninthauthor\unskip\ et~al.%
  }%
}
\def\aa@splitninth#1\and#2\aa@nil{\def\aa@ninthauthor{#1}}

\renewcommand*\aa@pageof{, page \thepage{} of \pageref*{LastPage}}
\makeatother

\usepackage[utf8]{inputenc}

\usepackage[switch, modulo]{lineno}
\nolinenumbers
\usepackage{euclid}
\nolinenumbers

\begin{document}
\nolinenumbers
%
%

\title{\Euclid}
\subtitle{Galaxy SED reconstruction in the PHZ processing function: impact on the PSF and the role of medium-band filters}  

\newcommand{\orcid}[1]{} 
\author{Euclid Collaboration: F.~Tarsitano\orcid{0000-0002-5919-0238}\thanks{\email{federica.tarsitano@phys.ethz.ch}}\inst{\ref{aff1},\ref{aff2}}
\and C.~Schreiber\inst{\ref{aff3}}
\and H.~Miyatake\orcid{0000-0001-7964-9766}\inst{\ref{aff4},\ref{aff5},\ref{aff6}}
\and A.~J.~Nishizawa\orcid{0000-0002-6109-2397}\inst{\ref{aff7},\ref{aff4}}
\and W.~G.~Hartley\inst{\ref{aff2}}
\and L.~Miller\orcid{0000-0002-3376-6200}\inst{\ref{aff8}}
\and C.~Cragg\inst{\ref{aff8},\ref{aff9}}
\and B.~Csizi\orcid{0000-0003-3227-6581}\inst{\ref{aff10}}
\and H.~Hildebrandt\orcid{0000-0002-9814-3338}\inst{\ref{aff11}}
\and B.~Altieri\orcid{0000-0003-3936-0284}\inst{\ref{aff12}}
\and A.~Amara\inst{\ref{aff13}}
\and S.~Andreon\inst{\ref{aff14}}
\and N.~Auricchio\orcid{0000-0003-4444-8651}\inst{\ref{aff15}}
\and C.~Baccigalupi\orcid{0000-0002-8211-1630}\inst{\ref{aff16},\ref{aff17},\ref{aff18},\ref{aff19}}
\and M.~Baldi\orcid{0000-0003-4145-1943}\inst{\ref{aff20},\ref{aff15},\ref{aff21}}
\and A.~Balestra\orcid{0000-0002-6967-261X}\inst{\ref{aff22}}
\and S.~Bardelli\orcid{0000-0002-8900-0298}\inst{\ref{aff15}}
\and A.~Biviano\orcid{0000-0002-0857-0732}\inst{\ref{aff17},\ref{aff16}}
\and E.~Branchini\orcid{0000-0002-0808-6908}\inst{\ref{aff23},\ref{aff24},\ref{aff14}}
\and M.~Brescia\orcid{0000-0001-9506-5680}\inst{\ref{aff25},\ref{aff26}}
\and J.~Brinchmann\orcid{0000-0003-4359-8797}\inst{\ref{aff27},\ref{aff28},\ref{aff29}}
\and S.~Camera\orcid{0000-0003-3399-3574}\inst{\ref{aff30},\ref{aff31},\ref{aff32}}
\and G.~Ca\~nas-Herrera\orcid{0000-0003-2796-2149}\inst{\ref{aff33},\ref{aff34}}
\and V.~Capobianco\orcid{0000-0002-3309-7692}\inst{\ref{aff32}}
\and C.~Carbone\orcid{0000-0003-0125-3563}\inst{\ref{aff35}}
\and V.~F.~Cardone\inst{\ref{aff36},\ref{aff37}}
\and J.~Carretero\orcid{0000-0002-3130-0204}\inst{\ref{aff38},\ref{aff39}}
\and S.~Casas\orcid{0000-0002-4751-5138}\inst{\ref{aff40},\ref{aff41}}
\and F.~J.~Castander\orcid{0000-0001-7316-4573}\inst{\ref{aff42},\ref{aff43}}
\and M.~Castellano\orcid{0000-0001-9875-8263}\inst{\ref{aff36}}
\and G.~Castignani\orcid{0000-0001-6831-0687}\inst{\ref{aff15}}
\and S.~Cavuoti\orcid{0000-0002-3787-4196}\inst{\ref{aff26},\ref{aff44}}
\and K.~C.~Chambers\orcid{0000-0001-6965-7789}\inst{\ref{aff45}}
\and A.~Cimatti\inst{\ref{aff46}}
\and C.~Colodro-Conde\inst{\ref{aff47}}
\and G.~Congedo\orcid{0000-0003-2508-0046}\inst{\ref{aff48}}
\and C.~J.~Conselice\orcid{0000-0003-1949-7638}\inst{\ref{aff49}}
\and L.~Conversi\orcid{0000-0002-6710-8476}\inst{\ref{aff50},\ref{aff12}}
\and Y.~Copin\orcid{0000-0002-5317-7518}\inst{\ref{aff51}}
\and F.~Courbin\orcid{0000-0003-0758-6510}\inst{\ref{aff52},\ref{aff53},\ref{aff54}}
\and H.~M.~Courtois\orcid{0000-0003-0509-1776}\inst{\ref{aff55}}
\and M.~Cropper\orcid{0000-0003-4571-9468}\inst{\ref{aff56}}
\and H.~Degaudenzi\orcid{0000-0002-5887-6799}\inst{\ref{aff2}}
\and G.~De~Lucia\orcid{0000-0002-6220-9104}\inst{\ref{aff17}}
\and H.~Dole\orcid{0000-0002-9767-3839}\inst{\ref{aff57}}
\and F.~Dubath\orcid{0000-0002-6533-2810}\inst{\ref{aff2}}
\and C.~A.~J.~Duncan\orcid{0009-0003-3573-0791}\inst{\ref{aff48}}
\and X.~Dupac\inst{\ref{aff12}}
\and S.~Escoffier\orcid{0000-0002-2847-7498}\inst{\ref{aff58}}
\and M.~Farina\orcid{0000-0002-3089-7846}\inst{\ref{aff59}}
\and R.~Farinelli\inst{\ref{aff15}}
\and S.~Farrens\orcid{0000-0002-9594-9387}\inst{\ref{aff60}}
\and F.~Faustini\orcid{0000-0001-6274-5145}\inst{\ref{aff36},\ref{aff61}}
\and S.~Ferriol\inst{\ref{aff51}}
\and F.~Finelli\orcid{0000-0002-6694-3269}\inst{\ref{aff15},\ref{aff62}}
\and N.~Fourmanoit\orcid{0009-0005-6816-6925}\inst{\ref{aff58}}
\and M.~Frailis\orcid{0000-0002-7400-2135}\inst{\ref{aff17}}
\and E.~Franceschi\orcid{0000-0002-0585-6591}\inst{\ref{aff15}}
\and M.~Fumana\orcid{0000-0001-6787-5950}\inst{\ref{aff35}}
\and S.~Galeotta\orcid{0000-0002-3748-5115}\inst{\ref{aff17}}
\and K.~George\orcid{0000-0002-1734-8455}\inst{\ref{aff63}}
\and W.~Gillard\orcid{0000-0003-4744-9748}\inst{\ref{aff58}}
\and B.~Gillis\orcid{0000-0002-4478-1270}\inst{\ref{aff48}}
\and C.~Giocoli\orcid{0000-0002-9590-7961}\inst{\ref{aff15},\ref{aff21}}
\and J.~Gracia-Carpio\inst{\ref{aff64}}
\and A.~Grazian\orcid{0000-0002-5688-0663}\inst{\ref{aff22}}
\and F.~Grupp\inst{\ref{aff64},\ref{aff65}}
\and S.~V.~H.~Haugan\orcid{0000-0001-9648-7260}\inst{\ref{aff66}}
\and H.~Hoekstra\orcid{0000-0002-0641-3231}\inst{\ref{aff34}}
\and W.~Holmes\inst{\ref{aff67}}
\and F.~Hormuth\inst{\ref{aff68}}
\and A.~Hornstrup\orcid{0000-0002-3363-0936}\inst{\ref{aff69},\ref{aff70}}
\and K.~Jahnke\orcid{0000-0003-3804-2137}\inst{\ref{aff71}}
\and M.~Jhabvala\inst{\ref{aff72}}
\and B.~Joachimi\orcid{0000-0001-7494-1303}\inst{\ref{aff9}}
\and E.~Keih\"anen\orcid{0000-0003-1804-7715}\inst{\ref{aff73}}
\and S.~Kermiche\orcid{0000-0002-0302-5735}\inst{\ref{aff58}}
\and A.~Kiessling\orcid{0000-0002-2590-1273}\inst{\ref{aff67}}
\and B.~Kubik\orcid{0009-0006-5823-4880}\inst{\ref{aff51}}
\and M.~K\"ummel\orcid{0000-0003-2791-2117}\inst{\ref{aff65}}
\and M.~Kunz\orcid{0000-0002-3052-7394}\inst{\ref{aff74}}
\and H.~Kurki-Suonio\orcid{0000-0002-4618-3063}\inst{\ref{aff75},\ref{aff76}}
\and R.~Laureijs\inst{\ref{aff77}}
\and A.~M.~C.~Le~Brun\orcid{0000-0002-0936-4594}\inst{\ref{aff78}}
\and S.~Ligori\orcid{0000-0003-4172-4606}\inst{\ref{aff32}}
\and P.~B.~Lilje\orcid{0000-0003-4324-7794}\inst{\ref{aff66}}
\and V.~Lindholm\orcid{0000-0003-2317-5471}\inst{\ref{aff75},\ref{aff76}}
\and I.~Lloro\orcid{0000-0001-5966-1434}\inst{\ref{aff79}}
\and G.~Mainetti\orcid{0000-0003-2384-2377}\inst{\ref{aff80}}
\and D.~Maino\inst{\ref{aff81},\ref{aff35},\ref{aff82}}
\and E.~Maiorano\orcid{0000-0003-2593-4355}\inst{\ref{aff15}}
\and O.~Mansutti\orcid{0000-0001-5758-4658}\inst{\ref{aff17}}
\and S.~Marcin\inst{\ref{aff83}}
\and O.~Marggraf\orcid{0000-0001-7242-3852}\inst{\ref{aff84}}
\and M.~Martinelli\orcid{0000-0002-6943-7732}\inst{\ref{aff36},\ref{aff37}}
\and N.~Martinet\orcid{0000-0003-2786-7790}\inst{\ref{aff85}}
\and F.~Marulli\orcid{0000-0002-8850-0303}\inst{\ref{aff86},\ref{aff15},\ref{aff21}}
\and R.~J.~Massey\orcid{0000-0002-6085-3780}\inst{\ref{aff87}}
\and E.~Medinaceli\orcid{0000-0002-4040-7783}\inst{\ref{aff15}}
\and S.~Mei\orcid{0000-0002-2849-559X}\inst{\ref{aff88},\ref{aff89}}
\and Y.~Mellier\thanks{Deceased}\inst{\ref{aff90},\ref{aff91}}
\and M.~Meneghetti\orcid{0000-0003-1225-7084}\inst{\ref{aff15},\ref{aff21}}
\and E.~Merlin\orcid{0000-0001-6870-8900}\inst{\ref{aff36}}
\and G.~Meylan\inst{\ref{aff92}}
\and A.~Mora\orcid{0000-0002-1922-8529}\inst{\ref{aff93}}
\and M.~Moresco\orcid{0000-0002-7616-7136}\inst{\ref{aff86},\ref{aff15}}
\and L.~Moscardini\orcid{0000-0002-3473-6716}\inst{\ref{aff86},\ref{aff15},\ref{aff21}}
\and R.~Nakajima\orcid{0009-0009-1213-7040}\inst{\ref{aff84}}
\and C.~Neissner\orcid{0000-0001-8524-4968}\inst{\ref{aff94},\ref{aff39}}
\and S.-M.~Niemi\orcid{0009-0005-0247-0086}\inst{\ref{aff33}}
\and C.~Padilla\orcid{0000-0001-7951-0166}\inst{\ref{aff94}}
\and F.~Pasian\orcid{0000-0002-4869-3227}\inst{\ref{aff17}}
\and J.~A.~Peacock\orcid{0000-0002-1168-8299}\inst{\ref{aff48}}
\and K.~Pedersen\inst{\ref{aff95}}
\and V.~Pettorino\orcid{0000-0002-4203-9320}\inst{\ref{aff33}}
\and S.~Pires\orcid{0000-0002-0249-2104}\inst{\ref{aff60}}
\and G.~Polenta\orcid{0000-0003-4067-9196}\inst{\ref{aff61}}
\and M.~Poncet\inst{\ref{aff96}}
\and L.~A.~Popa\inst{\ref{aff97}}
\and L.~Pozzetti\orcid{0000-0001-7085-0412}\inst{\ref{aff15}}
\and F.~Raison\orcid{0000-0002-7819-6918}\inst{\ref{aff64}}
\and A.~Renzi\orcid{0000-0001-9856-1970}\inst{\ref{aff98},\ref{aff99}}
\and J.~Rhodes\orcid{0000-0002-4485-8549}\inst{\ref{aff67}}
\and G.~Riccio\inst{\ref{aff26}}
\and E.~Romelli\orcid{0000-0003-3069-9222}\inst{\ref{aff17}}
\and M.~Roncarelli\orcid{0000-0001-9587-7822}\inst{\ref{aff15}}
\and C.~Rosset\orcid{0000-0003-0286-2192}\inst{\ref{aff88}}
\and R.~Saglia\orcid{0000-0003-0378-7032}\inst{\ref{aff65},\ref{aff64}}
\and Z.~Sakr\orcid{0000-0002-4823-3757}\inst{\ref{aff100},\ref{aff101},\ref{aff102}}
\and A.~G.~S\'anchez\orcid{0000-0003-1198-831X}\inst{\ref{aff64}}
\and D.~Sapone\orcid{0000-0001-7089-4503}\inst{\ref{aff103}}
\and B.~Sartoris\orcid{0000-0003-1337-5269}\inst{\ref{aff65},\ref{aff17}}
\and P.~Schneider\orcid{0000-0001-8561-2679}\inst{\ref{aff84}}
\and T.~Schrabback\orcid{0000-0002-6987-7834}\inst{\ref{aff10}}
\and M.~Scodeggio\inst{\ref{aff35}}
\and A.~Secroun\orcid{0000-0003-0505-3710}\inst{\ref{aff58}}
\and G.~Seidel\orcid{0000-0003-2907-353X}\inst{\ref{aff71}}
\and S.~Serrano\orcid{0000-0002-0211-2861}\inst{\ref{aff43},\ref{aff104},\ref{aff42}}
\and P.~Simon\inst{\ref{aff84}}
\and C.~Sirignano\orcid{0000-0002-0995-7146}\inst{\ref{aff98},\ref{aff99}}
\and G.~Sirri\orcid{0000-0003-2626-2853}\inst{\ref{aff21}}
\and L.~Stanco\orcid{0000-0002-9706-5104}\inst{\ref{aff99}}
\and J.~Steinwagner\orcid{0000-0001-7443-1047}\inst{\ref{aff64}}
\and P.~Tallada-Cresp\'{i}\orcid{0000-0002-1336-8328}\inst{\ref{aff38},\ref{aff39}}
\and A.~N.~Taylor\inst{\ref{aff48}}
\and H.~I.~Teplitz\orcid{0000-0002-7064-5424}\inst{\ref{aff105}}
\and I.~Tereno\orcid{0000-0002-4537-6218}\inst{\ref{aff106},\ref{aff107}}
\and N.~Tessore\orcid{0000-0002-9696-7931}\inst{\ref{aff56}}
\and S.~Toft\orcid{0000-0003-3631-7176}\inst{\ref{aff108},\ref{aff109}}
\and R.~Toledo-Moreo\orcid{0000-0002-2997-4859}\inst{\ref{aff110}}
\and F.~Torradeflot\orcid{0000-0003-1160-1517}\inst{\ref{aff39},\ref{aff38}}
\and I.~Tutusaus\orcid{0000-0002-3199-0399}\inst{\ref{aff42},\ref{aff43},\ref{aff101}}
\and J.~Valiviita\orcid{0000-0001-6225-3693}\inst{\ref{aff75},\ref{aff76}}
\and T.~Vassallo\orcid{0000-0001-6512-6358}\inst{\ref{aff17}}
\and A.~Veropalumbo\orcid{0000-0003-2387-1194}\inst{\ref{aff14},\ref{aff24},\ref{aff23}}
\and Y.~Wang\orcid{0000-0002-4749-2984}\inst{\ref{aff111}}
\and J.~Weller\orcid{0000-0002-8282-2010}\inst{\ref{aff65},\ref{aff64}}
\and G.~Zamorani\orcid{0000-0002-2318-301X}\inst{\ref{aff15}}
\and F.~M.~Zerbi\inst{\ref{aff14}}
\and I.~A.~Zinchenko\orcid{0000-0002-2944-2449}\inst{\ref{aff112}}
\and E.~Zucca\orcid{0000-0002-5845-8132}\inst{\ref{aff15}}
\and V.~Allevato\orcid{0000-0001-7232-5152}\inst{\ref{aff26}}
\and M.~Ballardini\orcid{0000-0003-4481-3559}\inst{\ref{aff113},\ref{aff114},\ref{aff15}}
\and M.~Bolzonella\orcid{0000-0003-3278-4607}\inst{\ref{aff15}}
\and E.~Bozzo\orcid{0000-0002-8201-1525}\inst{\ref{aff2}}
\and C.~Burigana\orcid{0000-0002-3005-5796}\inst{\ref{aff115},\ref{aff62}}
\and R.~Cabanac\orcid{0000-0001-6679-2600}\inst{\ref{aff101}}
\and A.~Cappi\inst{\ref{aff116},\ref{aff15}}
\and T.~Castro\orcid{0000-0002-6292-3228}\inst{\ref{aff17},\ref{aff18},\ref{aff16},\ref{aff117}}
\and J.~A.~Escartin~Vigo\inst{\ref{aff64}}
\and L.~Gabarra\orcid{0000-0002-8486-8856}\inst{\ref{aff8}}
\and J.~Garc\'ia-Bellido\orcid{0000-0002-9370-8360}\inst{\ref{aff118}}
\and S.~Hemmati\orcid{0000-0003-2226-5395}\inst{\ref{aff111}}
\and R.~Maoli\orcid{0000-0002-6065-3025}\inst{\ref{aff119},\ref{aff36}}
\and J.~Mart\'{i}n-Fleitas\orcid{0000-0002-8594-569X}\inst{\ref{aff120}}
\and M.~Maturi\orcid{0000-0002-3517-2422}\inst{\ref{aff100},\ref{aff121}}
\and N.~Mauri\orcid{0000-0001-8196-1548}\inst{\ref{aff46},\ref{aff21}}
\and R.~B.~Metcalf\orcid{0000-0003-3167-2574}\inst{\ref{aff86},\ref{aff15}}
\and N.~Morisset\inst{\ref{aff2}}
\and A.~Pezzotta\orcid{0000-0003-0726-2268}\inst{\ref{aff14}}
\and M.~P\"ontinen\orcid{0000-0001-5442-2530}\inst{\ref{aff75}}
\and I.~Risso\orcid{0000-0003-2525-7761}\inst{\ref{aff14},\ref{aff24}}
\and V.~Scottez\orcid{0009-0008-3864-940X}\inst{\ref{aff90},\ref{aff122}}
\and M.~Sereno\orcid{0000-0003-0302-0325}\inst{\ref{aff15},\ref{aff21}}
\and M.~Tenti\orcid{0000-0002-4254-5901}\inst{\ref{aff21}}
\and M.~Viel\orcid{0000-0002-2642-5707}\inst{\ref{aff16},\ref{aff17},\ref{aff19},\ref{aff18},\ref{aff117}}
\and M.~Wiesmann\orcid{0009-0000-8199-5860}\inst{\ref{aff66}}
\and Y.~Akrami\orcid{0000-0002-2407-7956}\inst{\ref{aff118},\ref{aff123}}
\and S.~Alvi\orcid{0000-0001-5779-8568}\inst{\ref{aff113}}
\and I.~T.~Andika\orcid{0000-0001-6102-9526}\inst{\ref{aff65}}
\and G.~Angora\orcid{0000-0002-0316-6562}\inst{\ref{aff26},\ref{aff113}}
\and S.~Anselmi\orcid{0000-0002-3579-9583}\inst{\ref{aff99},\ref{aff98},\ref{aff124}}
\and M.~Archidiacono\orcid{0000-0003-4952-9012}\inst{\ref{aff81},\ref{aff82}}
\and F.~Atrio-Barandela\orcid{0000-0002-2130-2513}\inst{\ref{aff125}}
\and L.~Bazzanini\orcid{0000-0003-0727-0137}\inst{\ref{aff113},\ref{aff15}}
\and D.~Bertacca\orcid{0000-0002-2490-7139}\inst{\ref{aff98},\ref{aff22},\ref{aff99}}
\and M.~Bethermin\orcid{0000-0002-3915-2015}\inst{\ref{aff126}}
\and A.~Blanchard\orcid{0000-0001-8555-9003}\inst{\ref{aff101}}
\and L.~Blot\orcid{0000-0002-9622-7167}\inst{\ref{aff127},\ref{aff78}}
\and M.~Bonici\orcid{0000-0002-8430-126X}\inst{\ref{aff128},\ref{aff35}}
\and S.~Borgani\orcid{0000-0001-6151-6439}\inst{\ref{aff129},\ref{aff16},\ref{aff17},\ref{aff18},\ref{aff117}}
\and M.~L.~Brown\orcid{0000-0002-0370-8077}\inst{\ref{aff49}}
\and S.~Bruton\orcid{0000-0002-6503-5218}\inst{\ref{aff130}}
\and A.~Calabro\orcid{0000-0003-2536-1614}\inst{\ref{aff36}}
\and B.~Camacho~Quevedo\orcid{0000-0002-8789-4232}\inst{\ref{aff16},\ref{aff19},\ref{aff17}}
\and F.~Caro\inst{\ref{aff36}}
\and C.~S.~Carvalho\inst{\ref{aff107}}
\and Y.~Charles\inst{\ref{aff85}}
\and F.~Cogato\orcid{0000-0003-4632-6113}\inst{\ref{aff86},\ref{aff15}}
\and S.~Conseil\orcid{0000-0002-3657-4191}\inst{\ref{aff51}}
\and A.~R.~Cooray\orcid{0000-0002-3892-0190}\inst{\ref{aff131}}
\and O.~Cucciati\orcid{0000-0002-9336-7551}\inst{\ref{aff15}}
\and S.~Davini\orcid{0000-0003-3269-1718}\inst{\ref{aff24}}
\and G.~Desprez\orcid{0000-0001-8325-1742}\inst{\ref{aff77}}
\and A.~D\'iaz-S\'anchez\orcid{0000-0003-0748-4768}\inst{\ref{aff132}}
\and S.~Di~Domizio\orcid{0000-0003-2863-5895}\inst{\ref{aff23},\ref{aff24}}
\and J.~M.~Diego\orcid{0000-0001-9065-3926}\inst{\ref{aff133}}
\and M.~Y.~Elkhashab\orcid{0000-0001-9306-2603}\inst{\ref{aff17},\ref{aff18},\ref{aff129},\ref{aff16}}
\and A.~Enia\orcid{0000-0002-0200-2857}\inst{\ref{aff15}}
\and Y.~Fang\orcid{0000-0002-0334-6950}\inst{\ref{aff65}}
\and A.~Finoguenov\orcid{0000-0002-4606-5403}\inst{\ref{aff75}}
\and A.~Franco\orcid{0000-0002-4761-366X}\inst{\ref{aff134},\ref{aff135},\ref{aff136}}
\and K.~Ganga\orcid{0000-0001-8159-8208}\inst{\ref{aff88}}
\and T.~Gasparetto\orcid{0000-0002-7913-4866}\inst{\ref{aff36}}
\and V.~Gautard\inst{\ref{aff137}}
\and E.~Gaztanaga\orcid{0000-0001-9632-0815}\inst{\ref{aff42},\ref{aff43},\ref{aff138}}
\and F.~Giacomini\orcid{0000-0002-3129-2814}\inst{\ref{aff21}}
\and F.~Gianotti\orcid{0000-0003-4666-119X}\inst{\ref{aff15}}
\and G.~Gozaliasl\orcid{0000-0002-0236-919X}\inst{\ref{aff139},\ref{aff75}}
\and M.~Guidi\orcid{0000-0001-9408-1101}\inst{\ref{aff20},\ref{aff15}}
\and C.~M.~Gutierrez\orcid{0000-0001-7854-783X}\inst{\ref{aff140}}
\and A.~Hall\orcid{0000-0002-3139-8651}\inst{\ref{aff48}}
\and C.~Hern\'andez-Monteagudo\orcid{0000-0001-5471-9166}\inst{\ref{aff141},\ref{aff47}}
\and J.~Hjorth\orcid{0000-0002-4571-2306}\inst{\ref{aff95}}
\and J.~J.~E.~Kajava\orcid{0000-0002-3010-8333}\inst{\ref{aff142},\ref{aff143}}
\and Y.~Kang\orcid{0009-0000-8588-7250}\inst{\ref{aff2}}
\and V.~Kansal\orcid{0000-0002-4008-6078}\inst{\ref{aff144},\ref{aff145}}
\and D.~Karagiannis\orcid{0000-0002-4927-0816}\inst{\ref{aff113},\ref{aff146}}
\and K.~Kiiveri\inst{\ref{aff73}}
\and J.~Kim\orcid{0000-0003-2776-2761}\inst{\ref{aff8}}
\and C.~C.~Kirkpatrick\inst{\ref{aff73}}
\and S.~Kruk\orcid{0000-0001-8010-8879}\inst{\ref{aff12}}
\and M.~Lattanzi\orcid{0000-0003-1059-2532}\inst{\ref{aff114}}
\and L.~Legrand\orcid{0000-0003-0610-5252}\inst{\ref{aff147},\ref{aff148}}
\and M.~Lembo\orcid{0000-0002-5271-5070}\inst{\ref{aff91},\ref{aff113},\ref{aff114}}
\and F.~Lepori\orcid{0009-0000-5061-7138}\inst{\ref{aff149}}
\and G.~Leroy\orcid{0009-0004-2523-4425}\inst{\ref{aff150},\ref{aff87}}
\and G.~F.~Lesci\orcid{0000-0002-4607-2830}\inst{\ref{aff86},\ref{aff15}}
\and J.~Lesgourgues\orcid{0000-0001-7627-353X}\inst{\ref{aff40}}
\and T.~I.~Liaudat\orcid{0000-0002-9104-314X}\inst{\ref{aff151}}
\and A.~Loureiro\orcid{0000-0002-4371-0876}\inst{\ref{aff152},\ref{aff153}}
\and J.~Macias-Perez\orcid{0000-0002-5385-2763}\inst{\ref{aff154}}
\and M.~Magliocchetti\orcid{0000-0001-9158-4838}\inst{\ref{aff59}}
\and C.~Mancini\orcid{0000-0002-4297-0561}\inst{\ref{aff35}}
\and F.~Mannucci\orcid{0000-0002-4803-2381}\inst{\ref{aff155}}
\and C.~J.~A.~P.~Martins\orcid{0000-0002-4886-9261}\inst{\ref{aff156},\ref{aff27}}
\and L.~Maurin\orcid{0000-0002-8406-0857}\inst{\ref{aff57}}
\and M.~Miluzio\inst{\ref{aff12},\ref{aff157}}
\and P.~Monaco\orcid{0000-0003-2083-7564}\inst{\ref{aff129},\ref{aff17},\ref{aff18},\ref{aff16}}
\and A.~Montoro\orcid{0000-0003-4730-8590}\inst{\ref{aff42},\ref{aff43}}
\and C.~Moretti\orcid{0000-0003-3314-8936}\inst{\ref{aff17},\ref{aff16},\ref{aff18}}
\and G.~Morgante\inst{\ref{aff15}}
\and S.~Nadathur\orcid{0000-0001-9070-3102}\inst{\ref{aff138}}
\and K.~Naidoo\orcid{0000-0002-9182-1802}\inst{\ref{aff138}}
\and P.~Natoli\orcid{0000-0003-0126-9100}\inst{\ref{aff113},\ref{aff114}}
\and A.~Navarro-Alsina\orcid{0000-0002-3173-2592}\inst{\ref{aff84}}
\and S.~Nesseris\orcid{0000-0002-0567-0324}\inst{\ref{aff118}}
\and M.~Oguri\orcid{0000-0003-3484-399X}\inst{\ref{aff158},\ref{aff159}}
\and L.~Pagano\orcid{0000-0003-1820-5998}\inst{\ref{aff113},\ref{aff114}}
\and D.~Paoletti\orcid{0000-0003-4761-6147}\inst{\ref{aff15},\ref{aff62}}
\and F.~Passalacqua\orcid{0000-0002-8606-4093}\inst{\ref{aff98},\ref{aff99}}
\and K.~Paterson\orcid{0000-0001-8340-3486}\inst{\ref{aff71}}
\and L.~Patrizii\inst{\ref{aff21}}
\and A.~Pisani\orcid{0000-0002-6146-4437}\inst{\ref{aff58}}
\and D.~Potter\orcid{0000-0002-0757-5195}\inst{\ref{aff149}}
\and G.~W.~Pratt\inst{\ref{aff60}}
\and S.~Quai\orcid{0000-0002-0449-8163}\inst{\ref{aff86},\ref{aff15}}
\and M.~Radovich\orcid{0000-0002-3585-866X}\inst{\ref{aff22}}
\and G.~Rodighiero\orcid{0000-0002-9415-2296}\inst{\ref{aff98},\ref{aff22}}
\and W.~Roster\orcid{0000-0002-9149-6528}\inst{\ref{aff64}}
\and S.~Sacquegna\orcid{0000-0002-8433-6630}\inst{\ref{aff160}}
\and M.~Sahl\'en\orcid{0000-0003-0973-4804}\inst{\ref{aff161}}
\and D.~B.~Sanders\orcid{0000-0002-1233-9998}\inst{\ref{aff45}}
\and E.~Sarpa\orcid{0000-0002-1256-655X}\inst{\ref{aff19},\ref{aff117},\ref{aff18}}
\and A.~Schneider\orcid{0000-0001-7055-8104}\inst{\ref{aff149}}
\and D.~Sciotti\orcid{0009-0008-4519-2620}\inst{\ref{aff36},\ref{aff37}}
\and E.~Sellentin\inst{\ref{aff162},\ref{aff34}}
\and L.~C.~Smith\orcid{0000-0002-3259-2771}\inst{\ref{aff163}}
\and J.~G.~Sorce\orcid{0000-0002-2307-2432}\inst{\ref{aff164},\ref{aff57}}
\and K.~Tanidis\orcid{0000-0001-9843-5130}\inst{\ref{aff8}}
\and C.~Tao\orcid{0000-0001-7961-8177}\inst{\ref{aff58}}
\and G.~Testera\inst{\ref{aff24}}
\and R.~Teyssier\orcid{0000-0001-7689-0933}\inst{\ref{aff165}}
\and S.~Tosi\orcid{0000-0002-7275-9193}\inst{\ref{aff23},\ref{aff24},\ref{aff14}}
\and A.~Troja\orcid{0000-0003-0239-4595}\inst{\ref{aff98},\ref{aff99}}
\and M.~Tucci\inst{\ref{aff2}}
\and A.~Venhola\orcid{0000-0001-6071-4564}\inst{\ref{aff166}}
\and D.~Vergani\orcid{0000-0003-0898-2216}\inst{\ref{aff15}}
\and G.~Verza\orcid{0000-0002-1886-8348}\inst{\ref{aff167},\ref{aff168}}
\and P.~Vielzeuf\orcid{0000-0003-2035-9339}\inst{\ref{aff58}}
\and S.~Vinciguerra\orcid{0009-0005-4018-3184}\inst{\ref{aff85}}
\and N.~A.~Walton\orcid{0000-0003-3983-8778}\inst{\ref{aff163}}
\and J.~R.~Weaver\orcid{0000-0003-1614-196X}\inst{\ref{aff169}}}
										   
\institute{Institute for Particle Physics and Astrophysics, Dept. of Physics, ETH Zurich, Wolfgang-Pauli-Strasse 27, 8093 Zurich, Switzerland\label{aff1}
\and
Department of Astronomy, University of Geneva, ch. d'Ecogia 16, 1290 Versoix, Switzerland\label{aff2}
\and
IBEX Innovations Ltd., NETPark Plexus, Thomas Wright Way, Sedgefield, TS21 3FD, UK\label{aff3}
\and
Kobayashi-Maskawa Institute for the Origin of Particles and the Universe, Nagoya University, Chikusa-ku, Nagoya, 464-8602, Japan\label{aff4}
\and
Institute for Advanced Research, Nagoya University, Chikusa-ku, Nagoya, 464-8601, Japan\label{aff5}
\and
Kavli Institute for the Physics and Mathematics of the Universe (WPI), University of Tokyo, Kashiwa, Chiba 277-8583, Japan\label{aff6}
\and
DX Center, Gifu Shotoku Gakuen University, 1-1 Takakuwanishi, Yanaizucho, Gifu, 501-6122, Japan\label{aff7}
\and
Department of Physics, Oxford University, Keble Road, Oxford OX1 3RH, UK\label{aff8}
\and
Department of Physics and Astronomy, University College London, Gower Street, London WC1E 6BT, UK\label{aff9}
\and
Universit\"at Innsbruck, Institut f\"ur Astro- und Teilchenphysik, Technikerstr. 25/8, 6020 Innsbruck, Austria\label{aff10}
\and
Ruhr University Bochum, Faculty of Physics and Astronomy, Astronomical Institute (AIRUB), German Centre for Cosmological Lensing (GCCL), 44780 Bochum, Germany\label{aff11}
\and
ESAC/ESA, Camino Bajo del Castillo, s/n., Urb. Villafranca del Castillo, 28692 Villanueva de la Ca\~nada, Madrid, Spain\label{aff12}
\and
School of Mathematics and Physics, University of Surrey, Guildford, Surrey, GU2 7XH, UK\label{aff13}
\and
INAF-Osservatorio Astronomico di Brera, Via Brera 28, 20122 Milano, Italy\label{aff14}
\and
INAF-Osservatorio di Astrofisica e Scienza dello Spazio di Bologna, Via Piero Gobetti 93/3, 40129 Bologna, Italy\label{aff15}
\and
IFPU, Institute for Fundamental Physics of the Universe, via Beirut 2, 34151 Trieste, Italy\label{aff16}
\and
INAF-Osservatorio Astronomico di Trieste, Via G. B. Tiepolo 11, 34143 Trieste, Italy\label{aff17}
\and
INFN, Sezione di Trieste, Via Valerio 2, 34127 Trieste TS, Italy\label{aff18}
\and
SISSA, International School for Advanced Studies, Via Bonomea 265, 34136 Trieste TS, Italy\label{aff19}
\and
Dipartimento di Fisica e Astronomia, Universit\`a di Bologna, Via Gobetti 93/2, 40129 Bologna, Italy\label{aff20}
\and
INFN-Sezione di Bologna, Viale Berti Pichat 6/2, 40127 Bologna, Italy\label{aff21}
\and
INAF-Osservatorio Astronomico di Padova, Via dell'Osservatorio 5, 35122 Padova, Italy\label{aff22}
\and
Dipartimento di Fisica, Universit\`a di Genova, Via Dodecaneso 33, 16146, Genova, Italy\label{aff23}
\and
INFN-Sezione di Genova, Via Dodecaneso 33, 16146, Genova, Italy\label{aff24}
\and
Department of Physics "E. Pancini", University Federico II, Via Cinthia 6, 80126, Napoli, Italy\label{aff25}
\and
INAF-Osservatorio Astronomico di Capodimonte, Via Moiariello 16, 80131 Napoli, Italy\label{aff26}
\and
Instituto de Astrof\'isica e Ci\^encias do Espa\c{c}o, Universidade do Porto, CAUP, Rua das Estrelas, PT4150-762 Porto, Portugal\label{aff27}
\and
Faculdade de Ci\^encias da Universidade do Porto, Rua do Campo de Alegre, 4150-007 Porto, Portugal\label{aff28}
\and
European Southern Observatory, Karl-Schwarzschild-Str.~2, 85748 Garching, Germany\label{aff29}
\and
Dipartimento di Fisica, Universit\`a degli Studi di Torino, Via P. Giuria 1, 10125 Torino, Italy\label{aff30}
\and
INFN-Sezione di Torino, Via P. Giuria 1, 10125 Torino, Italy\label{aff31}
\and
INAF-Osservatorio Astrofisico di Torino, Via Osservatorio 20, 10025 Pino Torinese (TO), Italy\label{aff32}
\and
European Space Agency/ESTEC, Keplerlaan 1, 2201 AZ Noordwijk, The Netherlands\label{aff33}
\and
Leiden Observatory, Leiden University, Einsteinweg 55, 2333 CC Leiden, The Netherlands\label{aff34}
\and
INAF-IASF Milano, Via Alfonso Corti 12, 20133 Milano, Italy\label{aff35}
\and
INAF-Osservatorio Astronomico di Roma, Via Frascati 33, 00078 Monteporzio Catone, Italy\label{aff36}
\and
INFN-Sezione di Roma, Piazzale Aldo Moro, 2 - c/o Dipartimento di Fisica, Edificio G. Marconi, 00185 Roma, Italy\label{aff37}
\and
Centro de Investigaciones Energ\'eticas, Medioambientales y Tecnol\'ogicas (CIEMAT), Avenida Complutense 40, 28040 Madrid, Spain\label{aff38}
\and
Port d'Informaci\'{o} Cient\'{i}fica, Campus UAB, C. Albareda s/n, 08193 Bellaterra (Barcelona), Spain\label{aff39}
\and
Institute for Theoretical Particle Physics and Cosmology (TTK), RWTH Aachen University, 52056 Aachen, Germany\label{aff40}
\and
Deutsches Zentrum f\"ur Luft- und Raumfahrt e. V. (DLR), Linder H\"ohe, 51147 K\"oln, Germany\label{aff41}
\and
Institute of Space Sciences (ICE, CSIC), Campus UAB, Carrer de Can Magrans, s/n, 08193 Barcelona, Spain\label{aff42}
\and
Institut d'Estudis Espacials de Catalunya (IEEC),  Edifici RDIT, Campus UPC, 08860 Castelldefels, Barcelona, Spain\label{aff43}
\and
INFN section of Naples, Via Cinthia 6, 80126, Napoli, Italy\label{aff44}
\and
Institute for Astronomy, University of Hawaii, 2680 Woodlawn Drive, Honolulu, HI 96822, USA\label{aff45}
\and
Dipartimento di Fisica e Astronomia "Augusto Righi" - Alma Mater Studiorum Universit\`a di Bologna, Viale Berti Pichat 6/2, 40127 Bologna, Italy\label{aff46}
\and
Instituto de Astrof\'{\i}sica de Canarias, E-38205 La Laguna, Tenerife, Spain\label{aff47}
\and
Institute for Astronomy, University of Edinburgh, Royal Observatory, Blackford Hill, Edinburgh EH9 3HJ, UK\label{aff48}
\and
Jodrell Bank Centre for Astrophysics, Department of Physics and Astronomy, University of Manchester, Oxford Road, Manchester M13 9PL, UK\label{aff49}
\and
European Space Agency/ESRIN, Largo Galileo Galilei 1, 00044 Frascati, Roma, Italy\label{aff50}
\and
Universit\'e Claude Bernard Lyon 1, CNRS/IN2P3, IP2I Lyon, UMR 5822, Villeurbanne, F-69100, France\label{aff51}
\and
Institut de Ci\`{e}ncies del Cosmos (ICCUB), Universitat de Barcelona (IEEC-UB), Mart\'{i} i Franqu\`{e}s 1, 08028 Barcelona, Spain\label{aff52}
\and
Instituci\'o Catalana de Recerca i Estudis Avan\c{c}ats (ICREA), Passeig de Llu\'{\i}s Companys 23, 08010 Barcelona, Spain\label{aff53}
\and
Institut de Ciencies de l'Espai (IEEC-CSIC), Campus UAB, Carrer de Can Magrans, s/n Cerdanyola del Vall\'es, 08193 Barcelona, Spain\label{aff54}
\and
UCB Lyon 1, CNRS/IN2P3, IUF, IP2I Lyon, 4 rue Enrico Fermi, 69622 Villeurbanne, France\label{aff55}
\and
Mullard Space Science Laboratory, University College London, Holmbury St Mary, Dorking, Surrey RH5 6NT, UK\label{aff56}
\and
Universit\'e Paris-Saclay, CNRS, Institut d'astrophysique spatiale, 91405, Orsay, France\label{aff57}
\and
Aix-Marseille Universit\'e, CNRS/IN2P3, CPPM, Marseille, France\label{aff58}
\and
INAF-Istituto di Astrofisica e Planetologia Spaziali, via del Fosso del Cavaliere, 100, 00100 Roma, Italy\label{aff59}
\and
Universit\'e Paris-Saclay, Universit\'e Paris Cit\'e, CEA, CNRS, AIM, 91191, Gif-sur-Yvette, France\label{aff60}
\and
Space Science Data Center, Italian Space Agency, via del Politecnico snc, 00133 Roma, Italy\label{aff61}
\and
INFN-Bologna, Via Irnerio 46, 40126 Bologna, Italy\label{aff62}
\and
University Observatory, LMU Faculty of Physics, Scheinerstr.~1, 81679 Munich, Germany\label{aff63}
\and
Max Planck Institute for Extraterrestrial Physics, Giessenbachstr. 1, 85748 Garching, Germany\label{aff64}
\and
Universit\"ats-Sternwarte M\"unchen, Fakult\"at f\"ur Physik, Ludwig-Maximilians-Universit\"at M\"unchen, Scheinerstr.~1, 81679 M\"unchen, Germany\label{aff65}
\and
Institute of Theoretical Astrophysics, University of Oslo, P.O. Box 1029 Blindern, 0315 Oslo, Norway\label{aff66}
\and
Jet Propulsion Laboratory, California Institute of Technology, 4800 Oak Grove Drive, Pasadena, CA, 91109, USA\label{aff67}
\and
Felix Hormuth Engineering, Goethestr. 17, 69181 Leimen, Germany\label{aff68}
\and
Technical University of Denmark, Elektrovej 327, 2800 Kgs. Lyngby, Denmark\label{aff69}
\and
Cosmic Dawn Center (DAWN), Denmark\label{aff70}
\and
Max-Planck-Institut f\"ur Astronomie, K\"onigstuhl 17, 69117 Heidelberg, Germany\label{aff71}
\and
NASA Goddard Space Flight Center, Greenbelt, MD 20771, USA\label{aff72}
\and
Department of Physics and Helsinki Institute of Physics, Gustaf H\"allstr\"omin katu 2, University of Helsinki, 00014 Helsinki, Finland\label{aff73}
\and
Universit\'e de Gen\`eve, D\'epartement de Physique Th\'eorique and Centre for Astroparticle Physics, 24 quai Ernest-Ansermet, CH-1211 Gen\`eve 4, Switzerland\label{aff74}
\and
Department of Physics, P.O. Box 64, University of Helsinki, 00014 Helsinki, Finland\label{aff75}
\and
Helsinki Institute of Physics, Gustaf H{\"a}llstr{\"o}min katu 2, University of Helsinki, 00014 Helsinki, Finland\label{aff76}
\and
Kapteyn Astronomical Institute, University of Groningen, PO Box 800, 9700 AV Groningen, The Netherlands\label{aff77}
\and
Laboratoire d'etude de l'Univers et des phenomenes eXtremes, Observatoire de Paris, Universit\'e PSL, Sorbonne Universit\'e, CNRS, 92190 Meudon, France\label{aff78}
\and
SKAO, Jodrell Bank, Lower Withington, Macclesfield SK11 9FT, UK\label{aff79}
\and
Centre de Calcul de l'IN2P3/CNRS, 21 avenue Pierre de Coubertin 69627 Villeurbanne Cedex, France\label{aff80}
\and
Dipartimento di Fisica "Aldo Pontremoli", Universit\`a degli Studi di Milano, Via Celoria 16, 20133 Milano, Italy\label{aff81}
\and
INFN-Sezione di Milano, Via Celoria 16, 20133 Milano, Italy\label{aff82}
\and
University of Applied Sciences and Arts of Northwestern Switzerland, School of Computer Science, 5210 Windisch, Switzerland\label{aff83}
\and
Universit\"at Bonn, Argelander-Institut f\"ur Astronomie, Auf dem H\"ugel 71, 53121 Bonn, Germany\label{aff84}
\and
Aix-Marseille Universit\'e, CNRS, CNES, LAM, Marseille, France\label{aff85}
\and
Dipartimento di Fisica e Astronomia "Augusto Righi" - Alma Mater Studiorum Universit\`a di Bologna, via Piero Gobetti 93/2, 40129 Bologna, Italy\label{aff86}
\and
Department of Physics, Institute for Computational Cosmology, Durham University, South Road, Durham, DH1 3LE, UK\label{aff87}
\and
Universit\'e Paris Cit\'e, CNRS, Astroparticule et Cosmologie, 75013 Paris, France\label{aff88}
\and
CNRS-UCB International Research Laboratory, Centre Pierre Bin\'etruy, IRL2007, CPB-IN2P3, Berkeley, USA\label{aff89}
\and
Institut d'Astrophysique de Paris, 98bis Boulevard Arago, 75014, Paris, France\label{aff90}
\and
Institut d'Astrophysique de Paris, UMR 7095, CNRS, and Sorbonne Universit\'e, 98 bis boulevard Arago, 75014 Paris, France\label{aff91}
\and
Institute of Physics, Laboratory of Astrophysics, Ecole Polytechnique F\'ed\'erale de Lausanne (EPFL), Observatoire de Sauverny, 1290 Versoix, Switzerland\label{aff92}
\and
Telespazio UK S.L. for European Space Agency (ESA), Camino bajo del Castillo, s/n, Urbanizacion Villafranca del Castillo, Villanueva de la Ca\~nada, 28692 Madrid, Spain\label{aff93}
\and
Institut de F\'{i}sica d'Altes Energies (IFAE), The Barcelona Institute of Science and Technology, Campus UAB, 08193 Bellaterra (Barcelona), Spain\label{aff94}
\and
DARK, Niels Bohr Institute, University of Copenhagen, Jagtvej 155, 2200 Copenhagen, Denmark\label{aff95}
\and
Centre National d'Etudes Spatiales -- Centre spatial de Toulouse, 18 avenue Edouard Belin, 31401 Toulouse Cedex 9, France\label{aff96}
\and
Institute of Space Science, Str. Atomistilor, nr. 409 M\u{a}gurele, Ilfov, 077125, Romania\label{aff97}
\and
Dipartimento di Fisica e Astronomia "G. Galilei", Universit\`a di Padova, Via Marzolo 8, 35131 Padova, Italy\label{aff98}
\and
INFN-Padova, Via Marzolo 8, 35131 Padova, Italy\label{aff99}
\and
Institut f\"ur Theoretische Physik, University of Heidelberg, Philosophenweg 16, 69120 Heidelberg, Germany\label{aff100}
\and
Institut de Recherche en Astrophysique et Plan\'etologie (IRAP), Universit\'e de Toulouse, CNRS, UPS, CNES, 14 Av. Edouard Belin, 31400 Toulouse, France\label{aff101}
\and
Universit\'e St Joseph; Faculty of Sciences, Beirut, Lebanon\label{aff102}
\and
Departamento de F\'isica, FCFM, Universidad de Chile, Blanco Encalada 2008, Santiago, Chile\label{aff103}
\and
Satlantis, University Science Park, Sede Bld 48940, Leioa-Bilbao, Spain\label{aff104}
\and
Infrared Processing and Analysis Center, California Institute of Technology, Pasadena, CA 91125, USA\label{aff105}
\and
Departamento de F\'isica, Faculdade de Ci\^encias, Universidade de Lisboa, Edif\'icio C8, Campo Grande, PT1749-016 Lisboa, Portugal\label{aff106}
\and
Instituto de Astrof\'isica e Ci\^encias do Espa\c{c}o, Faculdade de Ci\^encias, Universidade de Lisboa, Tapada da Ajuda, 1349-018 Lisboa, Portugal\label{aff107}
\and
Cosmic Dawn Center (DAWN)\label{aff108}
\and
Niels Bohr Institute, University of Copenhagen, Jagtvej 128, 2200 Copenhagen, Denmark\label{aff109}
\and
Universidad Polit\'ecnica de Cartagena, Departamento de Electr\'onica y Tecnolog\'ia de Computadoras,  Plaza del Hospital 1, 30202 Cartagena, Spain\label{aff110}
\and
Caltech/IPAC, 1200 E. California Blvd., Pasadena, CA 91125, USA\label{aff111}
\and
Astronomisches Rechen-Institut, Zentrum f\"ur Astronomie der Universit\"at Heidelberg, M\"onchhofstr. 12-14, 69120 Heidelberg, Germany\label{aff112}
\and
Dipartimento di Fisica e Scienze della Terra, Universit\`a degli Studi di Ferrara, Via Giuseppe Saragat 1, 44122 Ferrara, Italy\label{aff113}
\and
Istituto Nazionale di Fisica Nucleare, Sezione di Ferrara, Via Giuseppe Saragat 1, 44122 Ferrara, Italy\label{aff114}
\and
INAF, Istituto di Radioastronomia, Via Piero Gobetti 101, 40129 Bologna, Italy\label{aff115}
\and
Universit\'e C\^{o}te d'Azur, Observatoire de la C\^{o}te d'Azur, CNRS, Laboratoire Lagrange, Bd de l'Observatoire, CS 34229, 06304 Nice cedex 4, France\label{aff116}
\and
ICSC - Centro Nazionale di Ricerca in High Performance Computing, Big Data e Quantum Computing, Via Magnanelli 2, Bologna, Italy\label{aff117}
\and
Instituto de F\'isica Te\'orica UAM-CSIC, Campus de Cantoblanco, 28049 Madrid, Spain\label{aff118}
\and
Dipartimento di Fisica, Sapienza Universit\`a di Roma, Piazzale Aldo Moro 2, 00185 Roma, Italy\label{aff119}
\and
Aurora Technology for European Space Agency (ESA), Camino bajo del Castillo, s/n, Urbanizacion Villafranca del Castillo, Villanueva de la Ca\~nada, 28692 Madrid, Spain\label{aff120}
\and
Zentrum f\"ur Astronomie, Universit\"at Heidelberg, Philosophenweg 12, 69120 Heidelberg, Germany\label{aff121}
\and
ICL, Junia, Universit\'e Catholique de Lille, LITL, 59000 Lille, France\label{aff122}
\and
CERCA/ISO, Department of Physics, Case Western Reserve University, 10900 Euclid Avenue, Cleveland, OH 44106, USA\label{aff123}
\and
Laboratoire Univers et Th\'eorie, Observatoire de Paris, Universit\'e PSL, Universit\'e Paris Cit\'e, CNRS, 92190 Meudon, France\label{aff124}
\and
Departamento de F{\'\i}sica Fundamental. Universidad de Salamanca. Plaza de la Merced s/n. 37008 Salamanca, Spain\label{aff125}
\and
Universit\'e de Strasbourg, CNRS, Observatoire astronomique de Strasbourg, UMR 7550, 67000 Strasbourg, France\label{aff126}
\and
Center for Data-Driven Discovery, Kavli IPMU (WPI), UTIAS, The University of Tokyo, Kashiwa, Chiba 277-8583, Japan\label{aff127}
\and
Waterloo Centre for Astrophysics, University of Waterloo, Waterloo, Ontario N2L 3G1, Canada\label{aff128}
\and
Dipartimento di Fisica - Sezione di Astronomia, Universit\`a di Trieste, Via Tiepolo 11, 34131 Trieste, Italy\label{aff129}
\and
California Institute of Technology, 1200 E California Blvd, Pasadena, CA 91125, USA\label{aff130}
\and
Department of Physics \& Astronomy, University of California Irvine, Irvine CA 92697, USA\label{aff131}
\and
Departamento F\'isica Aplicada, Universidad Polit\'ecnica de Cartagena, Campus Muralla del Mar, 30202 Cartagena, Murcia, Spain\label{aff132}
\and
Instituto de F\'isica de Cantabria, Edificio Juan Jord\'a, Avenida de los Castros, 39005 Santander, Spain\label{aff133}
\and
INFN, Sezione di Lecce, Via per Arnesano, CP-193, 73100, Lecce, Italy\label{aff134}
\and
Department of Mathematics and Physics E. De Giorgi, University of Salento, Via per Arnesano, CP-I93, 73100, Lecce, Italy\label{aff135}
\and
INAF-Sezione di Lecce, c/o Dipartimento Matematica e Fisica, Via per Arnesano, 73100, Lecce, Italy\label{aff136}
\and
CEA Saclay, DFR/IRFU, Service d'Astrophysique, Bat. 709, 91191 Gif-sur-Yvette, France\label{aff137}
\and
Institute of Cosmology and Gravitation, University of Portsmouth, Portsmouth PO1 3FX, UK\label{aff138}
\and
Department of Computer Science, Aalto University, PO Box 15400, Espoo, FI-00 076, Finland\label{aff139}
\and
 Instituto de Astrof\'{\i}sica de Canarias, E-38205 La Laguna; Universidad de La Laguna, Dpto. Astrof\'\i sica, E-38206 La Laguna, Tenerife, Spain\label{aff140}
\and
Universidad de La Laguna, Dpto. Astrof\'\i sica, E-38206 La Laguna, Tenerife, Spain\label{aff141}
\and
Department of Physics and Astronomy, Vesilinnantie 5, University of Turku, 20014 Turku, Finland\label{aff142}
\and
Serco for European Space Agency (ESA), Camino bajo del Castillo, s/n, Urbanizacion Villafranca del Castillo, Villanueva de la Ca\~nada, 28692 Madrid, Spain\label{aff143}
\and
ARC Centre of Excellence for Dark Matter Particle Physics, Melbourne, Australia\label{aff144}
\and
Centre for Astrophysics \& Supercomputing, Swinburne University of Technology,  Hawthorn, Victoria 3122, Australia\label{aff145}
\and
Department of Physics and Astronomy, University of the Western Cape, Bellville, Cape Town, 7535, South Africa\label{aff146}
\and
DAMTP, Centre for Mathematical Sciences, Wilberforce Road, Cambridge CB3 0WA, UK\label{aff147}
\and
Kavli Institute for Cosmology Cambridge, Madingley Road, Cambridge, CB3 0HA, UK\label{aff148}
\and
Department of Astrophysics, University of Zurich, Winterthurerstrasse 190, 8057 Zurich, Switzerland\label{aff149}
\and
Department of Physics, Centre for Extragalactic Astronomy, Durham University, South Road, Durham, DH1 3LE, UK\label{aff150}
\and
IRFU, CEA, Universit\'e Paris-Saclay 91191 Gif-sur-Yvette Cedex, France\label{aff151}
\and
Oskar Klein Centre for Cosmoparticle Physics, Department of Physics, Stockholm University, Stockholm, SE-106 91, Sweden\label{aff152}
\and
Astrophysics Group, Blackett Laboratory, Imperial College London, London SW7 2AZ, UK\label{aff153}
\and
Univ. Grenoble Alpes, CNRS, Grenoble INP, LPSC-IN2P3, 53, Avenue des Martyrs, 38000, Grenoble, France\label{aff154}
\and
INAF-Osservatorio Astrofisico di Arcetri, Largo E. Fermi 5, 50125, Firenze, Italy\label{aff155}
\and
Centro de Astrof\'{\i}sica da Universidade do Porto, Rua das Estrelas, 4150-762 Porto, Portugal\label{aff156}
\and
HE Space for European Space Agency (ESA), Camino bajo del Castillo, s/n, Urbanizacion Villafranca del Castillo, Villanueva de la Ca\~nada, 28692 Madrid, Spain\label{aff157}
\and
Center for Frontier Science, Chiba University, 1-33 Yayoi-cho, Inage-ku, Chiba 263-8522, Japan\label{aff158}
\and
Department of Physics, Graduate School of Science, Chiba University, 1-33 Yayoi-Cho, Inage-Ku, Chiba 263-8522, Japan\label{aff159}
\and
INAF - Osservatorio Astronomico d'Abruzzo, Via Maggini, 64100, Teramo, Italy\label{aff160}
\and
Theoretical astrophysics, Department of Physics and Astronomy, Uppsala University, Box 516, 751 37 Uppsala, Sweden\label{aff161}
\and
Mathematical Institute, University of Leiden, Einsteinweg 55, 2333 CA Leiden, The Netherlands\label{aff162}
\and
Institute of Astronomy, University of Cambridge, Madingley Road, Cambridge CB3 0HA, UK\label{aff163}
\and
Univ. Lille, CNRS, Centrale Lille, UMR 9189 CRIStAL, 59000 Lille, France\label{aff164}
\and
Department of Astrophysical Sciences, Peyton Hall, Princeton University, Princeton, NJ 08544, USA\label{aff165}
\and
Space physics and astronomy research unit, University of Oulu, Pentti Kaiteran katu 1, FI-90014 Oulu, Finland\label{aff166}
\and
Center for Computational Astrophysics, Flatiron Institute, 162 5th Avenue, 10010, New York, NY, USA\label{aff167}
\and
International Centre for Theoretical Physics (ICTP), Strada Costiera 11, 34151 Trieste, Italy\label{aff168}
\and
MIT Kavli Institute for Astrophysics and Space Research, Massachusetts Institute of Technology, Cambridge, MA 02139, USA\label{aff169}}    





%
%
 \abstract{

Weak lensing surveys require accurate correction for the point spread function (PSF) when measuring galaxy shapes. For a diffraction-limited PSF, as arises in space-based missions, this correction depends on each galaxy’s spectral energy distribution (SED), hence a sufficiently accurate knowledge of the SED is needed for every galaxy. In the \Euclid mission, galaxy SED reconstruction, which is one of the tasks of the photometric-redshift processing function (PHZ PF), relies on broad- and medium-band ancillary photometry. The limited wavelength sampling of the \IE passband and signal-to-noise ratio may affect the reconstruction accuracy and translate into biases in the weak lensing measurements.
In this study, we present the methodology, which is employed in the \Euclid PHZ PF, for reconstructing galaxy SEDs at 55 wavelengths, sampling the \IE passband every 10 nm, and we assess whether it fulfils the accuracy requirements imposed on the \Euclid PSF model. 
We employ both physics- and data-driven methods, focusing on a new approach of template-based flux correction and Gaussian processes. For validation, we introduce an SED metric whose bias translates into inaccuracies in the PSF quadrupole moments.
Our findings demonstrate that Gaussian processes and template fitting meet the requirements only in specific, but complementary, redshift intervals. We therefore propose a hybrid approach, which leverages both methods. This solution proves to be effective in meeting the \Euclid accuracy requirements for most of the redshift range of the survey. Finally, we investigate the impact on the SED reconstruction of a new set of 16 evenly-spaced medium-band filters for the Subaru telescope, providing quasi-spectroscopic coverage of the \IE passband. This study shows promising results, ensuring accurate SED reconstruction and meeting the mission PSF requirements. This work thus provides not only the methodological foundation of galaxy SED reconstruction in the \Euclid PHZ PF, but also a roadmap for future improvements using a new medium-band survey.

}
%
%
\keywords{Galaxies: evolution, photometry; Cosmology; Methods: statistical, numerical.}
%
%
   \titlerunning{\Euclid Galaxy SED reconstruction}
   \authorrunning{Euclid
Collaboration: F. Tarsitano et al.}
   
   \maketitle

%
%
%
%
   
\section{Introduction}
\label{sec:S1}
   Our understanding of the Universe has been significantly changed by \cite{1998AJ....116.1009R} and \cite{1999ApJ...517..565P}, who provided observational evidence of an accelerated cosmic expansion. Since then, this discovery has been confirmed by a wide range of observations, yet the underlying theory is still debated, as competing explanations range from a change of the fundamental physics to generalisations of the FLRW cosmologies \citep[e.g.,][]{2006IJMPD..15.1753C, 2012Ap&SS.342..155B, 2012PhR...513....1C, 2017PhR...692....1N, 2018PhRvD..98b4038H, 2023PDU....3901163H}. Among the cosmological models proposed so far, the most relevant is $\Lambda$CDM, based on two elements, a positive cosmological constant ($\Lambda > 0$) and the cold dark matter (CDM) paradigm \citep{2022JHEAp..34...49A}. $\Lambda$CDM provides a simple yet powerful description of the Universe, based on parameters including the Hubble constant, which sets the expansion rate of the Universe; the relative densities of dark energy, dark matter and ordinary matter; the curvature of the Universe and other quantities describing the cosmic microwave background (CMB). Driven by advances in observational and computational capabilities, cosmologists have been making remarkable progress in constraining these parameters, with uncertainties of just a few percent, a precision that was unimaginable just three decades ago \citep{2022ARNPS..72....1T}.
    However, despite such advances, $\Lambda$CDM still lacks an adequate explanation for dark matter and dark energy \citep{2008ARA&A..46..385F, 2013PhR...530...87W, 2022NewAR..9501659P, 2025RSPTA.38340022E}. Advancing our current knowledge requires detailed measurements of both the expansion history of the Universe and the growth of structures as a function of cosmic time. This will be possible through Stage IV surveys such as \Euclid, a mission of the European Space Agency (ESA) whose core science is devoted to the investigation of the Dark Universe. 
   
   The Euclid Wide Survey (EWS; \citealt{Laureijs11, 2016SPIE.9904E..0OR, Scaramella-EP1, EuclidSkyOverview}) is set to cover 14\,000 deg$^2$ of the extragalactic sky in six years. It uses two instruments, VIS \citep{EuclidSkyVIS} and the Near-Infrared Spectrometer and Photometer \citep[NISP;][]{EuclidSkyNISP}. The VIS instrument is a broadband optical imager specifically designed for imaging galaxies in the wavelength range $550$--$900$\,nm, with a spatial resolution of 0\farcs18. NISP serves as an imager in the near-infrared (NIR) bands \YE, \JE, \HE, combining capabilities with a NIR slitless spectrograph \citep{Schirmer-EP18}. 
   
   \Euclid science focuses on two principal probes: galaxy clustering at $z>0.9$ through NIR slitless spectroscopy, and the matter distribution as a function of redshift using photometric galaxy clustering and weak gravitational lensing. The latter causes distortions of the images of distant galaxies whose light gets deflected by intervening large-scale structures. Due to this phenomenon, slight  distortions in the apparent shape of galaxies are observed, with adjacent galaxies exhibiting alignment patterns across the sky. Comprehensive reviews on the weak-lensing techniques were provided by \citet{2001PhR...340..291B}, \citet{2008ARNPS..58...99H}, \citet{2015RPPh...78h6901K} and \citet{2017SchpJ..1232440B}. Early studies were presented in \citet{2000Natur.405..143W}, using galaxy images observed with the Big Throughput Camera \citep{BTC}, in \citet{2000MNRAS.318..625B} using data from the William Herschel Telescope \citep{Bacon2001}, and by \citet{kaiser2000largescale} and \citet{2001A&A...374..757V}, providing shear measurements for the Canada France Hawaii Telescope survey \citep{1998yCat.1252....0M}. An analysis of CFHTLenS by \citet{2012MNRAS.427..146H, 2013MNRAS.432.2433H} was followed up with the first measurements from the Kilo-Degree Survey (KiDS; \citealt{KIDS2013, 2017MNRAS.465.1454H}). More recent works probed the cosmological model through cosmic shear measurements using the Dark Energy Survey 
   \citep{2018PhRvD..98d3528T, PhysRevD.105.023514, PhysRevD.105.023515, 2023PhRvD.107h3504A} and KiDS \citep{2021A&A...645A.104A, 2025A&A...703A.158W, 2025arXiv250319442S}.

   The major challenge in measuring the cosmic shear signal is that the typical change in a galaxy shape is tiny compared to other effects such as projection and statistical noise \citep{2015RPPh...78h6901K}. To cope with these relatively large statistical uncertainties, cosmic shear estimates are obtained by averaging over a large number of galaxies. During its observational campaign, \Euclid will provide shape measurements for billions of galaxies, allowing a significant decrease in statistical uncertainties through the averaging of individual measurements. However, for the results to be meaningful, and to fully exploit the potential of the \Euclid survey, a consistent reduction in the level of residual systematics is required. Instrumental effects altering galaxy shapes can overwhelm the lensing signal, and the most significant of them is the point spread function (PSF; \citealt{10.1111/j.1365-2966.2006.11315.x, 2005MNRAS.361.1287M}).

   In an effort to understand how PSF systematic errors affect weak lensing analyses, a series of studies presented analytical models propagating those errors through weak lensing measurements. \citet{2008A&A...484...67P} and \citet{2013MNRAS.429..661M}, hereafter M13, introduced an analytical framework assessing the multiplicative and additive biases in weak lensing shear that arise from errors in the assumed PSF model. Those authors showed that we expect multiplicative bias to be produced by errors in the assumed PSF size, and additive biases to be driven by errors in PSF size or ellipticity. The studies presented in M13 laid the foundation for a detailed propagation of systematic effects presented by \citet{2013MNRAS.431.3103C}, hereafter C13, where they focus on defining the characteristics and the requirements of space-based weak lensing experiments, and proposing error budgets for the various effects that enter a weak lensing analysis. This work, in turn, has been taken as a reference for the \Euclid mission. 

   Diffraction-limited telescopes, such as the \Euclid payload, have chromatic PSFs that depend on the SED of the star or galaxy being observed, due to the Fraunhofer diffraction pattern of the entrance pupil. Each wavelength in the \Euclid \IE bandpass makes a contribution to the integrated broadband PSF, where the monochromatic PSF contributions have sizes that increases with wavelength, but with more complex profile changes also present, arising from the mixture of optical wavefront error (which itself is chromatic), chromatic detector PSF contributions and achromatic guiding error. The effects of chromatic PSF models, in the context of the \Euclid mission, were studied by  \citet{2010MNRAS.405..494C}, hereafter C10, and \citet{2013MNRAS.432.2385S}. The latter focused on the bias introduced through the spatial variation of the galaxy colour. C10 explored a method to predict the PSF size as a function of galaxy colours, using stars with colours closer to the galaxies. An analogous work was conducted by \citet{2015ApJ...807..182M} for ground-based data. \citet{2018MNRAS.477.3433E}, hereafter EH18, revisited the method proposed in C10, studying the bias in the wavelength-dependent \Euclid PSF model caused by the limited broad-band-based information about the galaxy SEDs. Their findings suggested that the requirements indicated in M13 and C13 can be achieved using the photometric dataset used for \Euclid photometric redshift estimates. 
   
  Galaxy SEDs can be determined either through spectroscopy or photometry. In the first case, the emitted light is dispersed into its constituent wavelengths, allowing the measurement of the corresponding fluxes and the identification of spectral features such as emission and absorption lines. In the second case, the flux is measured through different filters, which can be narrow (typically in the range $[10,20]$\,\text{\AA}), medium (spanning the range $[100,200]$\,\text{\AA}) or broad (in the interval $[500,1000]$\,\text{\AA}). Multiple photometric observations are taken across a range of wavelengths, and the resulting data points are used to construct the SED. In a basic approach, the SED is given by interpolation methods (e.g., linear, cubic spline) of the flux points. Advanced regression methods are also available, such as the machine learning algorithm Gaussian processes (GP), which provide a functional form of the SED through a cross-correlation between flux points. While these methods can be used to reconstruct the galaxy SEDs solely relying on data, another approach is to make use of prior knowledge about the SED shapes. Known as template fitting (TF), this method compares the photometric sample to a set of predefined templates that represent different types of galaxies. Biases in this technique arise from potential template mismatches. 
  
  In comparison, spectroscopy appears to be ideally suited to the problem of SED reconstruction, but it is marred by several problems. Spectroscopy is usually less accurately calibrated than photometry. Spectra are also difficult to obtain for objects at the limiting magnitude of the \Euclid survey (AB limiting magnitude for $5 \sigma$ point-like sources are estimated to be 26.2 in \IE and 24.5 in \YE, \JE, and \HE, as reported in \cite{Scaramella-EP1}, and unfeasible for a sufficiently large number of galaxies to meet the aims of the mission.
  
  SED reconstruction represents a challenging task since galaxy SEDs encode important information on scales significantly smaller than the width of standard broad and medium-band filters. Therefore, a limited wavelength sampling of the \IE filter, combined with photometric inaccuracies, can lower the accuracy of the reconstruction.
  
  In the \Euclid survey, galaxy SEDs are one of the products of the photometric-redshift processing function (PHZ PF; \citealt{Q1-TP005}), a pipeline responsible for source photometric classification (into stars, galaxies, and quasars), photometric redshift estimates for the \Euclid core science, and galaxy physical properties for legacy science.   
  
  In this paper we conduct a study on galaxy SED reconstruction and its accuracy within the PHZ PF, using broad- and medium-band photometry present in the COSMOS field \citep{2007ApJS..172....1S}, which is part of the Euclid Auxiliary Fields (EAFs; \cite{Q1-TP001, EP-McPartland}). The EAFs (COSMOS, CDFS, SXDS, VVDS, AEGIS, GOODS-N) provide deep multi-wavelength data that are used to calibrate photometric redshifts, control the systematics in cosmic shear analyses, and characterise the galaxy sample of the EWS. In the context of this work, they constitute a sufficiently dense sampling of the $\IE$ passband to enable an accurate SED reconstruction. However, for most of the EWS area where only VIS and NISP photometry is available,  the medium-band information is missing. The methodology developed in this work is therefore designed for galaxies in the EAFs, using their broad- and medium- band photometry to reconstruct their SEDs. These then form the reference sample for the EWS, where the Nearest Neighbour Photometric Redshifts algorithm (NNPZ), described in \cite{Q1-TP005}, assigns to each galaxy a reconstructed SED obtained as the weighted mean of its 30 closest neighbours in flux space.
  

  In our work, we first investigate the optimal sampling of the \IE filter, and we compare the different reconstruction approaches mentioned above. Our goal is to establish a robust method to be used in PHZ PF for galaxy SED reconstruction, and to assess whether it achieves an accuracy level that is high enough to pass the requirements set for the \Euclid PSF model. Our work is therefore complementary to EH18: whereas they quantified feasibility, here we explore methodological strategies to minimise the bias in the SED, so that it does not exceed the accuracy level imposed on the PSF.
  

  As indicated in C13, the size and ellipticity of the chromatic PSF change with the SED of the source and the filter transmission as a function of wavelength. The corresponding error budget, defined as the uncertainty in the PSF model due to the wavelength dependence of the PSF (and therefore directly propagating from the inaccuracies on the SEDs) is estimated to be $b = 3.5\times 10^{-4}$. In the framework of this mission, an SED estimation that does not lead to PSF errors exceeding such a bias prevents a scenario in which dominant systematic effects are introduced in the PSF model, subsequently affecting cosmological inferences.
  
  Working with simulated galaxy SEDs, we start our analysis studying the impact that the scheme adopted to sample the \IE filter (defined according to how homogeneous and dense is the coverage of the passband) has on the SED reconstruction. Then we present a description of the metric we use to assess the accuracy of the SED reconstruction, followed by a detailed overview of GP and TF methods. We proceed by showing the results of the SED reconstruction through the available medium and broad-band photometry, comparing the different methodologies, and we discuss a combination of the approaches that minimises the bias in the SED metric. The SED reconstruction studied in this work is implemented in the \Euclid PHZ PF. 
  
  Finally, we present a new set of medium-band (MB) filters for the Subaru telescope, which ensures an evenly-spaced coverage of the \IE window, and we show that with this optimization it is possible to use a variety of reconstruction methods always meeting the \Euclid requirements. 

In this analysis, we adopt the \textit{Planck} 2018 cosmology \citep{2020A&A...641A...6P}: $H_0 = 67.4 \ \mathrm{km \ s^{-1} \ Mpc^{-1}}$, $\Omega_{\rm m} = 0.315$, and $\Omega_{\Lambda} = 0.685$.

\section{The SED metric}
\label{sec:S2}

In this section we assess how the sampling scheme of the \IE filter affects the accuracy we can achieve in the SED reconstruction. We calculate the fluxes at the sampled wavelengths by integrating the SEDs over idealised top-hat filters, and we interpolate the flux points to reconstruct the galaxy SEDs. We repeat this process at different sampling frequencies and assess the variations in accuracy. The latter is quantified through a metric that is able to translate the uncertainty on the SED reconstruction into bias in the PSF model.

\subsection{The \Euclid PSF model and the SED metric}
The \Euclid PSF model describes the PSF of a galaxy given its position, SED and telescope properties. As reported in \cite{EuclidSkyOverview}, it has been implemented as part of the \textit{Euclid PSF toolkit} which will be fully presented in a future publication (Duncan et al., in prep.).

In the \Euclid PSF model, the PSF of a galaxy, $\mathbf{P}_{\mathrm{gal}}(\mathbf{p})$, at position $\mathbf{p} = (x_{\mathrm{fov}}, y_{\mathrm{fov}})$ in the field of view, is computed as 
\begin{equation}
    \mathbf{P}_{\text{gal}}(\mathbf{p}) =
    \mathcal{O} \!\left[
      \frac{\int \diff \lambda \, T(\lambda, \mathbf{p}) \,\lambda S_{\text{gal}}(\lambda) \mathbf{P}(\lambda, \mathbf{p})}
           {\int \diff \lambda \, T(\lambda, \mathbf{p}) \,\lambda S_{\text{gal}}(\lambda)}
    \right],
	\label{eq:01}
\end{equation}
where $T(\lambda, \mathbf{p})$ is the throughput of the telescope and the \IE instrument at position $\mathbf{p}$, $\lambda S_{\text{gal}}(\lambda)$ denotes the galaxy SED expressed in photon counts per unit wavelength (up to a constant $1/hc$), and $\mathbf{P}(\lambda, \mathbf{p})$ is the monochromatic PSF at that position in the field of view. The operation $\mathcal{O}$ accounts for non-linear instrumental effects, such as the brighter-fatter effect and charge transfer inefficiencies. These effects, studied in \citet{Massey2015} and \citet{Israel2015}, are currently not included in the \Euclid PSF model and fall out of the scope of this work.

In \cref{eq:01}, the wavelength dependence of the PSF clearly illustrates the link between the galaxy SED and the PSF model. Furthermore, the role of the \Euclid PSF model is to provide values for $T$ and $\textbf{P}$, so that a PSF can be created for any given SED at any position in the field of view. The model numerically evaluates \cref{{eq:01}} integrating over a set of monochromatic PSFs. Indeed, our goal here is purely to reconstruct galaxy SEDs to limit the bias into the PSF. The resolution adopted for estimating the galaxy SED is discussed in \cref{sec:samplingVIS}.

For \Euclid, requirements on the PSF model are defined in terms of the PSF size and ellipticity metrics. These are characterised using quadrupole moments, which are second-order statistical measures of the PSF light distribution. The complex distortion effects influencing the size and shape of the PSF cause light to spread out far from the central peak so that, when integrating the light distribution of the PSF, the moments continue to increase or do not settle to a stable value. A standard solution to this problem is to apply a Gaussian weight function, $W_{xy}$, which emphasises the central peak of the PSF and attenuates the outer regions. Assuming that the centroid of the PSF is at the centre of the image $(x_0, y_0)$, we calculate the quadrupole moments as
\begin{align}
Q_{11} &= \frac{\sum_{xy} P_{xy} W_{xy} \,(x - x_0)^2}{\sum_{xy} P_{xy} W_{xy}} \ , \\
Q_{12} &= \frac{\sum_{xy} P_{xy} W_{xy} \,(x - x_0)(y - y_0)}{\sum_{xy} P_{xy} W_{xy}} \ , \\
Q_{22} &= \frac{\sum_{xy} P_{xy} W_{xy} \,(y - y_0)^2}{\sum_{xy} P_{xy} W_{xy}} \ .
\end{align}
They are used to infer the PSF size ($R^2_{\text{PSF}}$) and ellipticity (with components $\epsilon_1$ and $\epsilon_2$) as
\begin{equation}
R^2 = Q_{11} + Q_{22} \ , \quad
\epsilon_1 = \frac{Q_{11} - Q_{22}}{R^2} \ , \quad
\epsilon_2 = \frac{2 Q_{12}}{R^2} \ .
\label{eq:05}
\end{equation}

Quadrupole moments show a linear response to the PSF intensity. This implies that, when summing together multiple PSFs as in \cref{eq:01}, the moments get averaged and we can re-write the equations as
\begin{equation}
Q_{ij}(\mathbf{p}) =
\frac{\int \diff\lambda \, T(\lambda, \mathbf{p}) \, w(\lambda, \mathbf{p}) \, \lambda S_{\text{gal}}(\lambda) \, Q_{ij}(\lambda, \mathbf{p})}
     {\int \diff\lambda \, T(\lambda, \mathbf{p}) \, w(\lambda, \mathbf{p}) \, \lambda S_{\text{gal}}(\lambda)} \ .
\label{eq:06}
\end{equation}

In \cref{eq:06}, $Q_{ij}(\lambda, \mathbf{p})$ are the quadrupole moments of each monochromatic PSF and $w(\lambda, \mathbf{p}) < 1$ is a correction term accounting for the flux loss derived from the application of the Gaussian weight function to the PSF light distribution. By pre-computing these monochromatic quadrupole moments, the integral above can be evaluated for various choices of SED, $S_{\rm gal}(\lambda)$.

Using the same formalism as in \cref{eq:01}, we define the SED metric as
\begin{equation}
    \mathcal{M}_{\lambda} = 
    \frac{\int \lambda^{2} \, T(\lambda) \, S_{\text{gal}}(\lambda) \, \diff\lambda}
         {\int \lambda \, T(\lambda) \, S_{\text{gal}}(\lambda) \, \diff\lambda} \ ,
	\label{eq:07}
\end{equation}
where $T(\lambda)$ is the \IE transmission curve. This new metric can be interpreted as the effective central wavelength of the \IE pass-band for a given SED. Thanks to its integral form encoding the information used to build the \Euclid PSF, biases in this metric translate straightforwardly into biases in the PSF quadrupole
moments, and hence into the PSF size and ellipticity. In the end, the comparison of this metric calculated using simulated and reconstructed SEDs will allow us to test the accuracy in the SED reconstruction against the \Euclid PSF requirements.

\subsection{Sampling the VIS passband}
\label{sec:samplingVIS}
We consider a set of 10\,000 simulated galaxy SEDs (simulations are described in \cref{sec:S3}). 
The photometric dataset of each object is computed as a set of evenly spaced wavelengths by integrating the SED over juxtaposed and adjacent idealised top-hat filters. An interpolation of the flux points reconstructs the SED of the galaxy. This process is repeated across different sets of equidistant wavelengths, each time reducing the distance between flux points ($ \Delta\lambda$ hereafter), allowing us to assess the variations in accuracy on the SED reconstruction.
The accuracy estimates are obtained through comparison between the simulated (true) SED and the reconstructed one. 

Using \cref{eq:07} we compute the metric for both the simulated and the reconstructed SEDs, denoted by $\mathcal{M}_{\lambda}^{\mathrm{true}}$ and $\mathcal{M}_{\lambda}^{\mathrm{rec}}$. The bias is defined as the fractional error between them:
\begin{equation}
     b = \frac{\mathcal{M}_{\lambda}^\text{true} - \mathcal{M}_{\lambda}^\text{rec}}
              {\mathcal{M}_{\lambda}^\text{true}} \ .
    \label{eq:bias}
\end{equation}

\begin{figure}[tbp]
    \centering
   \includegraphics[width=\columnwidth]{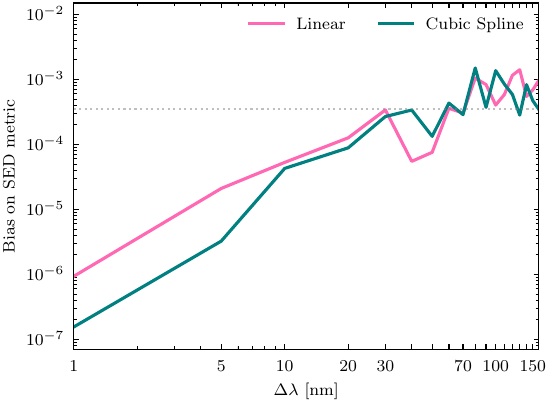}
   \caption{Median fractional bias in the SED metric arising from the interpolation of galaxy SEDs from coarse sampling down to 1\,nm sampling using linear (pink line) and cubic spline (dark green) interpolation. The dotted line represents the \Euclid requirement set on the PSF model.}
    \label{fig:01}
\end{figure}

For every $\mathrm{\Delta \lambda}$ we calculate the bias in the SED metric for each galaxy, obtaining an overall bias distribution, and we estimate the sampling accuracy through its median value. We assume a bias of $b = 3.5\times 10^{-4}$ as our target accuracy. This threshold is indicated in C13 as the budget error on shear measurements in the framework of the \Euclid mission, due to the wavelength-dependency of the PSF model. Our findings are reported in \cref{fig:01}. 
The displayed trends meet the requirement on the \Euclid PSF at $\mathrm{\Delta \lambda < 10 \ nm}$ and downwards, and show a progressive degradation from a binning of about 30\,nm and upwards. The increasing scatter observed for $\Delta\lambda > 30$\,nm arises from undersampling, failing to capture essential features of the SEDs, including emission lines, breaks, and spectral slopes, whose inclusion is critical in order to minimise errors in the reconstruction.

Our findings suggest that we should infer the galaxy SED at $55$ wavelengths evenly spaced every $10$\,nm between 450 and 1000\,nm. These results are obtained from an idealised test, where the galaxy SEDs are sampled through top-hat filters that provide a uniform and complete coverage of the \IE pass-band. When using real filter configurations (described in \cref{sec:S3}), which are non-uniformly distributed and present gaps between the filters, a larger bias is  expected. From \cref{fig:01} we see that a sampling of $\mathrm{\Delta \lambda < 10 \ nm}$ is needed to ensure that the bias remains smaller than $b = 3.5\times 10^{-4}$ when using a simple interpolation method. 

\section{Data}
\label{sec:S3}
The results we present in this paper are based on measurements computed from simulated galaxy SEDs. In order for the conclusions drawn from our work to be applicable to actual \Euclid analyses, we must ensure that these simulated data are sufficiently representative of the target galaxy population used for weak lensing cosmology. Among the \Euclid Photometric Redshift Auxiliary fields (PHZ AUX; see \citealt{EuclidSkyOverview}), the most suitable field that can be used to build a training reference sample for galaxy SED estimation is the COSMOS field. We therefore build our simulations to represent the wide range of photometric data available in this field.

Generation of our simulated galaxy SEDs comprises two steps: construction of a galaxy population model, and then sampling from this model, including the application of an appropriate amount of photometric noise to the derived fluxes. While realism in the simulated data is important, it is not necessary that we match the galaxy population exactly.

Furthermore, we target an optimistic, best-case scenario in which we can neglect potential systematic uncertainties arising from the calibration and photometric extraction steps of the overall \Euclid pipeline. Namely, we do not model the effects of photometric zeropoint errors, the impact of unknown wavelength variation in filter responses (see \citealt{EP-Paltani}) or biases in photometric measurement due to source blending. In this way, our results will reflect the degree of accuracy that can in principle be achieved with currently-available data and the SED reconstruction methods we employ, but in practice are an upper bound to performance.

\subsection{Galaxy population model}

A suitable galaxy population model will consist of the evolving abundance of galaxies with redshift and luminosity, together with a description of the SEDs of those galaxies. For the former, we use pre-existing measurements of the $i$-band galaxy luminosity function produced for various tests of the \Euclid photometric redshift pipeline, and computed using the COSMOS catalogue of \citet{2022ApJS..258...11W}. The COSMOS data were cut at $i<25.5$ and fit with a series of single Schechter functions \citep{Schechter1976} in narrow redshift intervals ($\Delta z=0.1$ at low $z$), widening at higher redshifts to account for the lower number density, and assuming for each galaxy that the best-fit photometric redshift given in the \citet{Laigle2016} catalogue is the true redshift. The standard ${\rm V}/{\rm V_{max}}$ \citep{Schmidt1968} estimator was used for the luminosity function and the three Schechter function parameters were then determined by the maximum a posteriori solution with flat priors in $\logten(\Phi^*)$, $M^*$ and $\alpha$. Modelling the galaxy luminosity function is not the subject of this paper, and so we refer the interested reader to \citet{Weigel2016} for an extensive description of methods for doing so.

The second aspect of the galaxy population model is a description of the diversity of galaxy rest-frame SEDs as a function of redshift and luminosity. In recent years a number of methods have been developed with the aim of being able to rapidly produce mock galaxy samples for tasks involving simulation-based inference \citep[e.g.][]{Herbel2017, Amara2021, Moser2024}. Typically, these methods are based upon a set of template SED components that can be linearly combined to form a total galaxy SED. An equal number of redshift-evolving Dirichlet parameters can then be fit to a known galaxy sample, which allows coefficients for these template components to be sampled when simulating a population. While these methods show great promise, they do not yet explicitly incorporate luminosity dependence in the model and so for our purposes we prefer a more direct data-driven approach. For internal consistency in our population model we use the same set of photometry that was used to measure redshifts and luminosities for the luminosity functions, namely the COSMOS catalogue of \citet{Laigle2016}.

\begin{figure}
    \centering
   \includegraphics[width=\columnwidth]{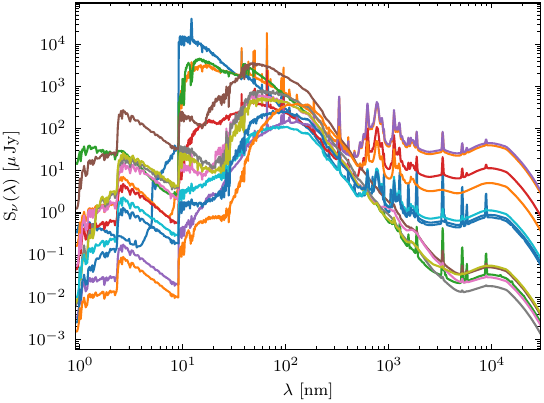}
   \caption{Set of baseline SED templates used to build the galaxy population model, displayed in the optical to infrared window. Each galaxy is simulated through a linear combination of the elements of this basis.}
    \label{fig:seds}
\end{figure}

To build the SED model, we use the \texttt{EAzY} photometric redshift code \citep{Brammer2008} to fit the set of twelve UltraVISTA \citep{{McCracken2012}} SED components that are provided by the \texttt{EAzY} package to each galaxy, fixing their redshifts to the same ones used in the luminosity function measurement. Each catalogued object therefore has a redshift, $i$-band luminosity, and a set of template coefficients that we use to construct its rest-frame SED. 
The SED of our population model then consists of drawing a galaxy within an appropriate range of redshift and luminosity, and using the set of coefficients for that galaxy to build the SED. The search ranges we use are $\pm 0.2$ in redshift and $ \pm 0.5 $ mag in luminosity. If no objects lie within the search criteria, then we drop the luminosity constraint and draw from the correct interval in redshift. The basis set of linearly interpolated SED templates is reproduced in \cref{fig:seds}.

\subsection{Simulations of galaxy SEDs}

With the galaxy population model in hand we proceed to the production of our simulated data set. We draw galaxies across the set of luminosity functions using the same cosmology as was assumed during their measurement, taking care to balance the number drawn from each redshift interval appropriately. We then assign an SED to each redshift and luminosity pair as described in the previous sub-section. We scale the amplitude of the SED such that it reflects the drawn value of $i$-band luminosity, redshift it and finally integrate it through the set of broad and medium-band passbands used in this study to obtain observed-frame noiseless fluxes.

The set of passbands we use reflects the data that are currently available in the COSMOS field, including \Euclid observations, and are a super-set of those described in \citet{Laigle2016}. These data will eventually be incorporated into the \Euclid ecosystem for use within the pipeline and are taken from a range of sources in addition to \Euclid: the DES Deep Fields \citep[DECam $ugrizY$,][]{Hartley2022}, HSC SSP \citep[HSC $grizy$,][]{Aihara2018,Aihara2022}, COSMOS SuprimeCam (SC) \citep[$BgVriz^{\prime}z^{\prime\prime}$ + 12 medium bands,][]{Capak2007,Taniguchi2007,Taniguchi2015}, CLAUDS \citep[CFHT $u$,][]{Sawicki2019}, UltraVISTA \citep[VISTA $YJHKs$,][]{McCracken2012} and the Cosmic Dawn Survey \citep[Spitzer IRAC ch1, ch2,][]{Moneti-EP17}. The central wavelengths, band widths, and depths ($5\sigma$, $2^{\arcsec}$-diameter apertures) of the various data sets are described in \cref{tab:filters}. Most of these data are included in the updated COSMOS catalogue of \citep{2022ApJS..258...11W} and we adopt their measured depths. In the case of UltraVISTA, the listed depths correspond to the ultra-deep stripes as further observations have been obtained to even out the depth across the field. DECam depths for the 3-year version of the DES Deep Fields are given in \citet{Hartley2022} and \Euclid depths reflect the requirements placed on the signal-to-noise ratio (S/N) of the PHZ AUX fields \citep{EuclidSkyOverview}. 

\begin{table}[ht]
\centering
\caption{Characteristics of the filters used in this work.}
\begin{tabularx}{\columnwidth}{Xrrr}
\toprule
Filter Name & Central $\lambda$ ($\AA$) & Bandwidth ($\AA$) & Depth \\
\midrule
$\mathrm{MegaCam \ CFHT} \ u$ & 3682 & 1200 & 26.50 \\
$\mathrm{DECam} \ u$ & 3876 & 570 & 25.64 \\
$\mathrm{SuprimeCam \ IA427}$ & 4263 & 207 & 24.80 \\
$\mathrm{SuprimeCam} \ B$ & 4454 & 892 & 26.50 \\
$\mathrm{SuprimeCam \ IA464}$ & 4635 & 218 & 24.30 \\
$\mathrm{SuprimeCam} \ g$ & 4771 & 1265 & 24.80 \\
$\mathrm{HSC} \ g$ & 4812 & 1500 & 26.80 \\
$\mathrm{DECam}  \ g$ & 4826 & 1480 & 26.46 \\
$\mathrm{SuprimeCam \ IA484}$ & 4849 & 229 & 25.20 \\
$\mathrm{SuprimeCam \ IA505}$ & 5062 & 231 & 24.80 \\
$\mathrm{SuprimeCam \ IA527}$ & 5261 & 243 & 25.10 \\
$\mathrm{SuprimeCam} \ V$ & 5464 & 1900 & 25.50 \\
$\mathrm{SuprimeCam \ IA574}$ & 5764 & 273 & 24.50 \\
$\mathrm{HSC} \ r$ & 6230 & 1547 & 26.50 \\
$\mathrm{SuprimeCam \ IA624}$ & 6232 & 300 & 25.10 \\
$\mathrm{SuprimeCam} \ r$ & 6274 & 1960 & 25.80 \\
$\mathrm{DECam} \ r$ & 6432 & 1480 & 25.73 \\
$\mathrm{SuprimeCam \ IA679}$ & 6780 & 336 & 24.30 \\
$\mathrm{SuprimeCam \ IA709}$ & 7075 & 316 & 24.60 \\
$\Euclid \ \mathrm{VIS} \ I_E$ & 7180 & 3900 & 25.80 \\
$\mathrm{SuprimeCam \ IA738}$ & 7360 & 324 & 24.80 \\
$\mathrm{SuprimeCam} \ i$ & 7667 & 2590 & 25.40 \\
$\mathrm{SuprimeCam \ IA767}$ & 7686 & 365 & 24.30 \\
$\mathrm{HSC} \ i$ & 7702 & 1471 & 26.30 \\
$\mathrm{DECam} \ i$ & 7826 & 1470 & 25.54 \\
$\mathrm{SuprimeCam \ IA827}$ & 8244 & 343 & 24.30 \\
$\mathrm{HSC} \ z$ & 8903 & 766 & 25.90 \\
$\mathrm{SuprimeCam \ \it z^{+}}$ & 9041 & 847 & 24.40 \\
$\mathrm{SuprimeCam \ \it z^{++}}$ & 9099 & 1335 & 25.00 \\
$\mathrm{DECam} \ z$ & 9178 & 1520 & 24.97 \\
$\mathrm{HSC} \ Y$ & 9771 & 1810 & 25.20 \\
$\mathrm{DECam}  \ Y$ & 9899 & 1530 & 23.70 \\
$\mathrm{UltraVISTA} \ Y$ & 10214 & 923 & 24.70 \\
$\Euclid \ \mathrm{NISP} \ I_Y$ & 10858 & 2630 & 24.25 \\
$\mathrm{UltraVISTA} \ J$ & 12535 & 1718 & 25.00 \\
$\Euclid \ \mathrm{NISP} \ I_J$ & 13685 & 4510 & 24.25 \\
$\mathrm{UltraVISTA} \ H$ & 16454 & 2905 & 24.70 \\
$\Euclid \ \mathrm{NISP} \ I_H$ & 17739 & 5670 & 24.25 \\
$\mathrm{UltraVISTA \ \it K_s}$ & 21540 & 3074 & 24.40 \\
\bottomrule
\end{tabularx}
\tablefoot{Central wavelengths and bandwidths are given in \AA, depths are 5$\sigma$ limiting magnitudes in 2\arcsec-diameter apertures.}
\label{tab:filters}
\end{table}

The vast majority of galaxies that will be used in \Euclid cosmology analyses are small on the sky, and the \Euclid photometric pipeline produces colours through PSF-matched aperture measurements. We therefore simulate photometric noise consistently, adopting a simplified flux uncertainty model. Gaussian scatter is applied to the fluxes based on the aperture depths in \cref{tab:filters}, with a constant aperture correction of $0.3$\,mag across all bands. This approach assumes perfect PSF-matching, and consequently does not capture the full S/N complexity, particularly for bright, spatially extended galaxies, where the S/N is likely to be overestimated. In this work, we do not attempt to replicate a detailed S/N–magnitude relation, as this is not expected to significantly impact our goal of exploring the upper limit performance of galaxy SED reconstruction methods. Finally, to account for the variations in depth level that are observed in most of the bands that are used for the SED reconstruction, we apply an object-by-object Gaussian scatter to the expected depths at the level of $0.1$–$0.3$\,mag, depending on the band. At constant noiseless apparent brightness, the simulated signal to noise will vary from object to object within a band. After noise is applied, we discard any galaxies for which $ \IE > 25.5 $, ensuring a sample that is sufficiently deep for our analysis and includes objects that could be up-scattered from fainter magnitudes in the EWS. We continue the simulation until we reach a sample of 60\,000 accepted objects.



\section{Flux and SED reconstruction}
\label{sec:S4}
In \cref{sec:S2} we described the tests conducted to assess the sampling scheme of the \IE filter that minimises the bias in the SED metric.
In order to attain this optimal scheme, we examine different techniques: (i) interpolation methods, such as linear and b-spline interpolation; (ii) GP, a machine-learning algorithm which is purely based on data; and (iii) a template-fitting based approach, which is driven by physics. We describe (ii) and (iii) in the following paragraphs, including the detailed techniques we adopt for SED reconstruction and bias estimation. 

\begin{figure}
   \centering
   \includegraphics[width=\columnwidth]{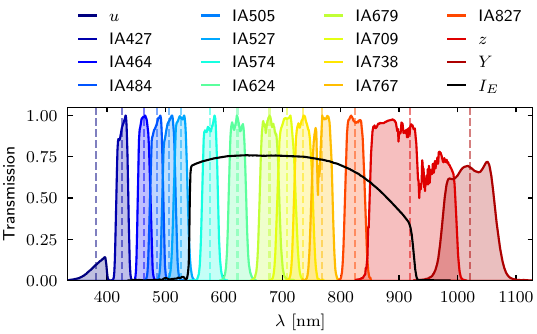}
   \caption{Coverage of the \IE filter obtained using the Subaru SC medium band filters, and the broad pass-bands $u$ and $z$ from DECam and $Y$ from VISTA VIRCam. The vertical dashed lines are placed at the weighted central wavelength of each filter.}
    \label{fig:02_altern}
\end{figure}

\subsection{Gaussian processes}
\label{sec:4.2}
Data-driven approaches process the photometric point fluxes in order to obtain a mathematical model that we can use to predict the fluxes at all wavelengths. These algorithms do not rely on SED templates. In this work we explore the performance of GP. We use the module \texttt{GaussianProcessRegressor} presented in \citet{10.7551/mitpress/3206.001.0001} and available in the \texttt{Python} library \texttt{Scikit-learn} \citep{scikit-learn}. GP is an unsupervised algorithm consisting of a main function and a covariance function, also known as \textit{kernel}, which quantifies the correlation between data points. The advantage of this method is that it does not assume a fixed mathematical model for regression, but it starts from a distribution of functions that is updated during the fitting process according to the observed data points and their inaccuracies. The inferred output is a function averaging over a posterior model distribution. The kernel determines the smoothness and linearity of the functions. We run initial tests with two different kernels: radial basis function (RBF) and Matérn. We found that the second provides a more accurate method for SED reconstruction, therefore we choose it for the rest of the analysis. 
We also investigated the impact of the kernel characteristic \textit{length scale}, which controls the correlation between flux points at different wavelength separations and which can therefore affect the smoothness of the output function. We tested both fixed length-scale values, representative of the average separation between flux points, as well as the configuration where the algorithm adaptively optimises this parameter, and we found no significant impact on the reconstruction accuracy.
For more details about the regression with GP and the properties of the two kernels, and their hyperparameters, we refer the interested reader to \cref{sec:AppGP}. 
We ran GP over a set of filters including the Subaru SC medium bands, the DECam \textit{u} and \textit{z}, and the VIRCAM \textit{Y} broad pass-bands (\cref{fig:02_altern}). DECam \textit{u} and the VIRCAM \textit{Y} are needed to constrain the SED outside of the \IE band. In a first exploratory phase of the analysis, we tested GP using different combinations of these bands, with the purpose of studying how the algorithm behaves at the bluest and reddest ends of the SED-reconstruction interval. We found that omitting either DECam \textit{u} or VIRCAM \textit{Y} leads to boundary effects, as the algorithm lacks anchor points at the edges and tends to extrapolate the SED incorrectly. Including both filters mitigates these artefacts and stabilises the reconstruction within the \IE\ passband. DECam \textit{z} is also included because it partially falls within the \IE wavelength range. We exclude the other broad-bands as they tend to smooth over the SED. 
The output of the GP can be evaluated at any wavelength together with uncertainties.

\begin{figure*}
\includegraphics[width=\textwidth]{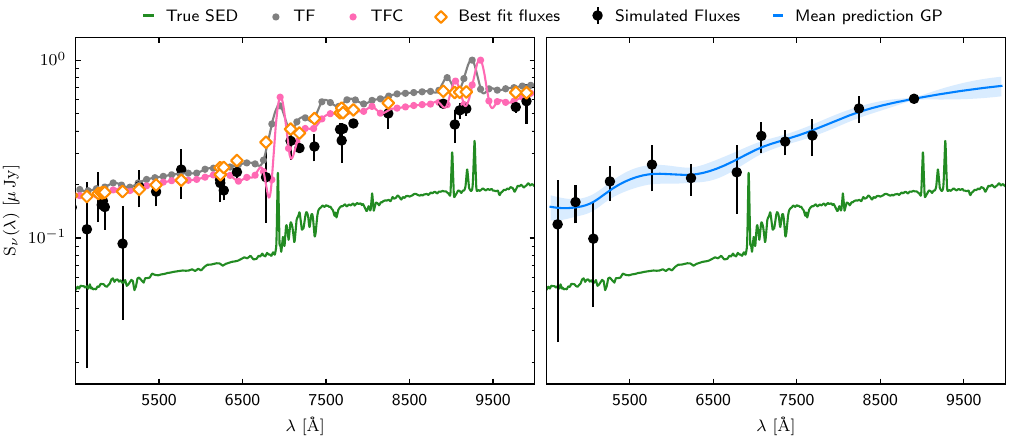}
\caption{Example of SED reconstruction for a galaxy randomly drawn from the simulated sample. The true SED is plotted in green. \textit{Left panel:} example of SED reconstruction using template-based flux correction (TFC). The black points are the true fluxes known from simulations, and the orange-contoured white diamonds represent their best-fit estimates. The grey line shows the SED reconstructed with TF only, and the pink line is the colour correction. True and reconstructed SEDs are shifted in flux to enhance readability. \textit{Right panel:} example of SED reconstruction using GP regression starting from the subset of medium-band simulated fluxes.
The blue line (with the respective $95\%$ confidence intervals) shows the inferred galaxy SED using the Matérn kernel.}
\label{fig:04}
\end{figure*}

\subsection{Template-based flux correction}
\label{sec:4.1}
TF compares the photometry of each object with a library of SEDs. The latter is a set of predefined templates, representing different types of galaxies. 
For TF we use the \texttt{Phosphoros} package (Paltani et al., in prep). The novelty of \texttt{Phosphoros} is to offer a fully Bayesian model that has versatile prior distributions for all parameters. \texttt{Phosphoros} has already been used in the \Euclid photo-$z$ challenge \citep{Desprez-EP10}, whose aim was to study the precision of different techniques for photometric redshift estimation against the requirements imposed by \Euclid cosmic-shear analyses. \texttt{Phosphoros} has been validated and compared to a similar code, \texttt{Le Phare} \citep{2006A&A...457..841I, 2011ascl.soft08009A} in \citet{2023A&A...670A..82D}.
\texttt{Phosphoros} assigns to each input galaxy a multivariate posterior distribution including photo-$z$, reddening ($E_{B-V}$), SED-index, and luminosity. The best model from the posterior distribution provides an estimate of the fluxes at any wavelength.
While the templates are usually considered sufficient to determine photometric redshifts, their small number means that they are unable to accurately infer the true SEDs, so that the reconstructed fluxes will be significantly biased. In order to mitigate this, we introduce a template-based flux correction (TFC) method, where the estimated fluxes are corrected using their neighbouring fluxes. 

In the TFC approach, for each reconstructed flux we first consider its neighbouring simulated fluxes, denoted as$f_\mathrm{L}^{\mathrm{true}}$ (the flux in the preceding filter, on the left) and $f_\mathrm{R}^{\mathrm{true}}$ (the flux in the following filter, on the right). The best estimates of $f_\mathrm{L}^{\mathrm{true}}$ and $f_\mathrm{R}^{\mathrm{true}}$ are reconstructed through \texttt{Phosphoros} and denoted as $f_\mathrm{L}^{\mathrm{Ph}}$ and $f_\mathrm{R}^{\mathrm{Ph}}$, respectively. Then on each side, we assume that the ratio between the true flux and the corrected one is equal to the ratio between the corresponding fluxes reconstructed through TF. This colour-corrected flux can be written as
\begin{equation}
    \frac{f_\mathrm{L}^{\mathrm{true}}}{f_\mathrm{L}^{*}} = \frac{f_\mathrm{L}^\mathrm{Ph}}{f^{*,\mathrm{Ph}}}, \qquad 
    \frac{f_\mathrm{R}^{\mathrm{true}}}{f_\mathrm{R}^{*}} = \frac{f_\mathrm{R}^\mathrm{Ph}}{f^{*,\mathrm{Ph}}} \, ,
\label{eq:eq_CS_1}
\end{equation}
where $f^{*,\mathrm{Ph}}$ is one of the 55 fluxes inferred through TF. Also, $f_\mathrm{L}^{*}$ and $f_\mathrm{R}^{*}$ are its colour-corrected values. We derive them as
\begin{equation}
    f_\mathrm{L}^{*} = \frac{f_\mathrm{L}^{\mathrm{true}}}{f_\mathrm{L}^\mathrm{Ph}} f^{*,\mathrm{Ph}}, \qquad 
    f_\mathrm{R}^{*} = \frac{f_\mathrm{R}^{\mathrm{true}}}{f_\mathrm{R}^\mathrm{Ph}} f^{*,\mathrm{Ph}} \, .
\label{eq:eq_CS_2}
\end{equation}
We further manipulate $f_\mathrm{L}^{*}$ and $f_\mathrm{R}^{*}$ to get a final estimate for the flux. In this last step we take in to account the distance that separates it from its neighbours on the left ($d_\mathrm{L}$) and on the right ($d_\mathrm{R}$). The weighted estimated flux is calculated as
\begin{equation}
    f^{*} = \frac{d_\mathrm{L} f_\mathrm{L}^{*} + d_\mathrm{R} f_\mathrm{R}^{*}}{d_\mathrm{L} + d_\mathrm{R}} \, .
\label{eq:eq_CS_3}
\end{equation}

This method serves to weaken the model dependence of the results, which are obtained using an idealised discrete set of templates which tends to under-represent the variety of features we observe in real galaxy SEDs.

\cref{tab:filters} describes the set of filters employed in TF. \Cref{fig:04} displays the SED of a simulated galaxy, the true fluxes and their estimates from TF, and the fluxes reconstructed with TF at the 55 chosen wavelengths. The latter get a correction in colour space.

\subsection{SED reconstruction}
\label{sec:S5}
\Cref{fig:04} shows an example of reconstructed SED for a galaxy randomly drawn from the simulated dataset. To evaluate the accuracy, we calculate the metric for both the true and the inferred galaxy SED and estimate the bias, as explained in \cref{sec:S2}. Next, we divide the galaxy sample into redshift bins, using the photometric redshift determined through TF with \texttt{Phosphoros}. We target redshifts up to $z=3$ and set a bin width of $0.2$. For each galaxy falling in a certain redshift bin, we calculate the bias in the SED metric, obtaining a bias distribution for each bin. By averaging the values in each distribution, we assess how the accuracy in SED reconstruction varies as a function of redshift. To avoid constraining the galaxy sample to specific redshift intervals, we shift the redshift bins and recompute the average SED metric using the galaxies in the new bins. The results of this analysis are presented in the next section.



\section{Results}
\label{sec:S6}
\begin{figure}
   \centering
   \includegraphics[width=\columnwidth]{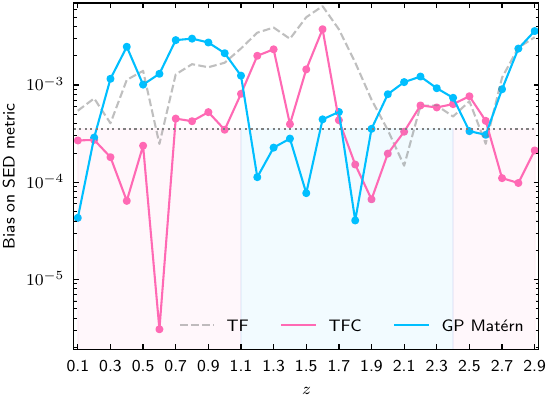}
   \caption{Bias in SED metric as a function of photometric redshift, calculated using different methods for SED reconstruction: colour-space corrected TF (TFC; magenta)  and GP (blue). The plot displays the bias averaged over single redshift bins. The dashed grey line represents the results obtained with TF before colour correction. The dotted horizontal line highlights the contribution of the PSF wavelength-dependency to the bias accounted for the \Euclid PSF model. }
    \label{fig:05}
\end{figure}

\begin{figure}
   \centering
   \includegraphics[width=\columnwidth]{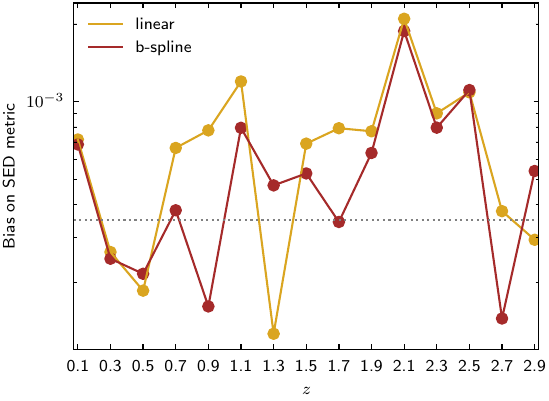}
   \caption{Bias in the SED metric averaged over single redshift bins, calculated from the SED reconstruction using linear (sand-coloured) and b-spline (in brown) interpolation, respectively.}
    \label{fig:06}
\end{figure}

Following the methods for galaxy SED reconstruction outlined above, we study the bias in the SED metric as a function of photometric redshift. The results from TFC and GP are shown in \cref{fig:05}. We observe that the different reconstruction methods meet the \Euclid requirements only in certain redshift intervals, which differ between approaches. More specifically, our findings indicate that GP can ensure SED reconstruction below the bias requirement in the redshift interval $[1.1, 1.8]$, with lower accuracy in the other ranges, where -- however -- TFC proves to be more robust. \Cref{fig:06} shows that linear and b-spline interpolation methods, which perform well at low redshift, do not maintain the same accuracy of TFC and GP in the remaining redshift intervals. We also quantify the dispersion of the individual bias in the SED metric in each redshift bin, finding an average value of $0.001$ for TFC, $0.005$ for GP, and $0.015$ for the interpolation methods. Therefore, TFC and GP yield lower dispersion in the bias in the SED metric bias compared to interpolation methods, proving their robustness in the reconstruction.

The results suggest the use of a combined method for reconstructing galaxy SEDs. This approach involves both TFC and GP, applied to different photometric redshift intervals. In this way it is possible to perform SED reconstruction, using existing photometric data of the COSMOS field, with an accuracy that passes the requirements imposed by the \Euclid mission PSF model in the majority of tomographic bins. 

\section{Discussion}
\label{sec:S7}

\subsection{Performance of SED reconstruction methods}
Our findings demonstrate that a careful analysis of the bias introduced in the reconstruction of galaxy SEDs can highlight the strengths of various adopted methods, allowing us to combine them and optimise the accuracy achieved in the reconstruction itself. The combination of TFC and GP is shown to be capable of meeting the requirements imposed on a mission like \Euclid, as defined in C13, except for a limited amount of redshift bins where the bias trend fluctuates close to the tolerance level. 

GP regression is used to infer the overall galaxy SED using the available medium-band photometry. In this approach, features that are typical of galaxy SEDs are smoothed out. This is due to the nature of GP, which is purely data-driven and does not follow specific prescriptions given by the underlying physics of galaxies. More precisely, the H$\alpha$ emission line falls within the \IE filter up to redshift $z \simeq 0.37$, H$\beta$ is contained in the interval $z\in [0.13,0.85]$, and \ion{O}{iii} in $z\in [0.10,0.80]$. The Balmer break lies within the \IE\ window for window for $0.51 \lesssim z \lesssim 1.47$.
The inability of GP to capture such SED sharp features can potentially introduce errors in the SED reconstruction, which explains their poor performance up to $z\simeq 1.1$. In the redshift interval [1.1, 1.8] spectral features become broadened or are poorly sampled by the available photometry. TFC, which relies on matching the latter to a discrete set of templates through the presence of identifiable spectral features, may therefore suffer from template mismatch, leading to a degradation of the reconstruction accuracy. By contrast, GP tend to smooth the trend, resulting in a better performance where the continuum dominates or sharp features do not provide strong constraints. 
To summarise, TFC and GP show stability in mutually exclusive $z$ bins and can be used to complement each other.

\subsection{Galaxy SED reconstruction in the \Euclid PHZ PF}
\label{sec:phzpf}
The results presented in this work have been directly implemented in \Euclid PHZ PF, where galaxy SEDs are part of the products delivered to enable both \Euclid core and legacy science. The strategy adopted within the pipeline is to recover galaxy SEDs as a set of flux points evenly spaced every 10\,nm in the \IE passband, corresponding to 55 narrow-band top-hat idealised filters. This choice originates from the analysis presented in \cref{sec:samplingVIS}, which shows that coarser sampling schemes would introduce systematic effects exceeding the constraints imposed on the \Euclid PSF. 

SEDs are reconstructed within the PHZ PF using the NNPZ algorithm, which recovers information on \Euclid galaxies combining a reference sample with a nearest-neighbour approach (more details in \citealt{Q1-TP005}). Reference samples used to infer star SEDs are built from \textit{Gaia} BP/RP spectrophotometry matched to \Euclid sources, assuming that \textit{Gaia} fully covers all stellar types and metallicities across the \Euclid footprint. Narrow-band fluxes in the 55 filters are computed by integrating the \textit{Gaia} spectra. A similar approach is not feasible for galaxy SEDs, as spectrophotometric datasets are incomplete and fail to match \Euclid's depth. Therefore, reference samples are drawn from special fields like COSMOS, and galaxy SEDs are reconstructed following the methodology developed in this work, combining TFC and GP. 

This implementation ensures that the PHZ PF produces galaxy SEDs within the accuracy level required for chromatic PSF in most $z$ bins. The methodological framework established here has thus become a key component of the \Euclid PHZ pipeline, building upon the previous feasibility studies on the \Euclid chromatic PSF and serving an operational requirement of the mission.

\subsection{Galaxy SEDs with a new set of HSC medium-band filters}
\label{sec:MBnew}

\begin{figure}
   \centering
   \includegraphics[width=\columnwidth]{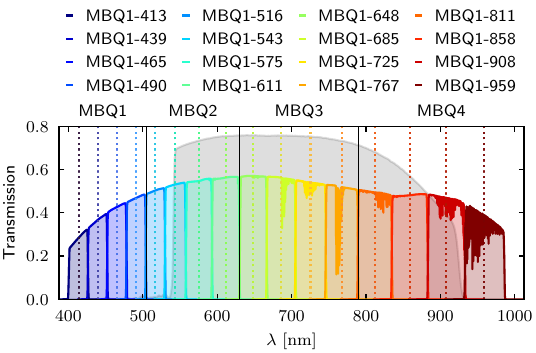}
   \caption{
   The new HSC medium-band filter design, with 16 filters organised in four quadrants (MBQ1--4). Each filter is named after its weighted central wavelength, indicated by a dashed vertical line. The \IE filter is shown in the background.}
    \label{fig:07}
\end{figure}

As stressed above, the SED reconstruction presented so far is limited by the COSMOS dataset, which is uneven across the \IE filter, and provides only a poor coverage towards the red end of the passband. In this section we describe the results obtained running our reconstruction pipeline on galaxies simulated from a new set of filters. More precisely, we forecast the accuracy we could reach if galaxy SEDs were reconstructed starting from a newly fabricated set of 16  Subaru/Hyper Suprime-Cam (HSC) medium-band filters spaced every $\sim 30$\,nm and evenly distributed across the \IE window\footnote{For more details we refer the reader to \url{https://subarutelescope.org//Observing/Instruments/HSC/sensitivity.html}.}. The MB are organised in four quadrants (MBQ1--4), with the first (MBQ1) already commissioned\footnote{Detailed information on MBQ1 can be found at \url{https://subarutelescope.org/Instruments/HSC/fig/HSC_mbq1.png}.}. This new filter design is displayed in \cref{fig:07}. 
The filter transmission curves include the following effects:
\begin{itemize}
    \item the quantum efficiency of the CCD, which is the fraction of photons that the CCD can convert into electronic signal;
    \item the dewar window sealing the CCD to reduce thermal noise. It can absorb or reflect a small portion of light, therefore affecting the transmission;
    \item the transmittance of the Primary Focus Unit of the HSC, quantifying how much light is allowed through the optics;
    \item the reflectivity of the primary mirror of the Subaru telescope, measuring the fraction of light reflected by the mirror at different wavelengths;
    \item the atmospheric absorption at Mauna Kea.
\end{itemize}
A catalogue of simulated photometry (for $\mathrm{\sim 1h}$ exposure time) and synthetic spectra, based on the EL COSMOS DR1 release \citep{EL_COSMOS2020}, serves as baseline for our interpolation and regression methods for SED reconstruction. \cref{tab:MBfilters} summarises the properties of each filter.

\begin{figure}[h!]
   \centering
   \includegraphics[width=\columnwidth]{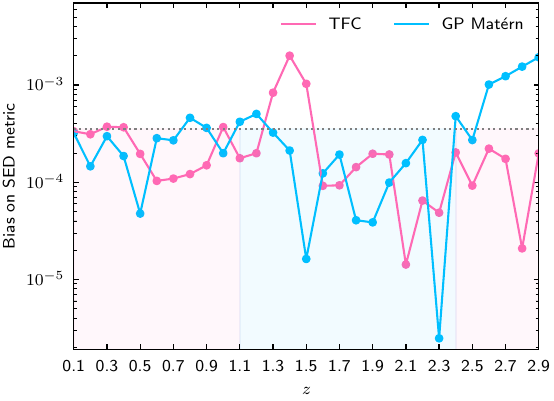}
   \caption{Bias in the SED metric calculated on galaxy SEDs reconstructed through a photometric dataset based on the filter design proposed in \cref{fig:07}. Magenta and brown lines display the accuracy reached through TFC and GP, respectively.}
    \label{fig:08}
\end{figure}

\begin{figure}[h!]
   \centering
   \includegraphics[width=\columnwidth]{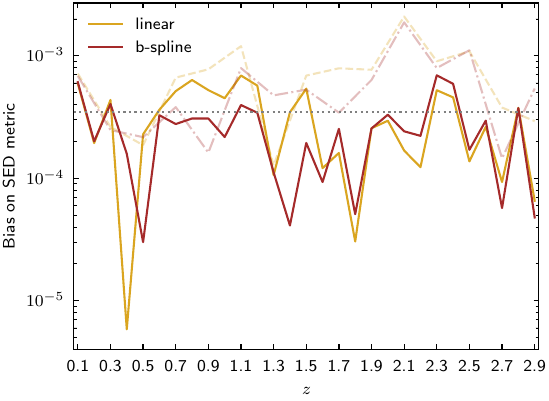}
   \caption{Bias in the SED metric calculated using linear and b-spline interpolation, coloured in sand and brown, respectively. The bias is averaged over a distribution of uncertainties per redshfit bin. Galaxy SEDs were reconstructed through a photometric dataset based on the filter design proposed in \cref{fig:07}. For comparison, the results from \cref{fig:06} are also shown, with the dashed sand-coloured line representing the linear interpolation, and the dash-dotted brown line the b-spline interpolation.}
    \label{fig:11}
\end{figure}

\begin{table*}[h!]
\centering
\caption{Characteristics of the new HSC-MB survey with the Subaru telescope.}
\label{tab:MBfilters}
\renewcommand{\arraystretch}{1.2}

\begin{tabular*}{\textwidth}{@{\extracolsep{\fill}}lcccccc}
\toprule
Filter Name & Central $\lambda$ [nm] & Bandwidth [nm] 
& \multicolumn{3}{c}{Limiting magnitude [mag]} \\
\cmidrule(lr){4-6}
& & & $T_{\rm exp}=5$ min & $T_{\rm exp}=10$ min & $T_{\rm exp}=60$ min \\
\midrule
 \  MB413 & 413.34 & 25.87 & 25.04 & 25.43 & 26.41 \\
 \  MB439 & 439.20 & 25.87 & 25.14 & 25.52 & 26.51 \\
 \  MB465 & 465.10 & 25.82 & 25.09 & 25.47 & 26.45 \\
 \  MB490 & 490.98 & 25.87 & 25.33 & 25.71 & 26.69 \\
 \  MB516 & 516.86 & 25.87 & 25.57 & 25.95 & 26.93 \\
 \  MB543 & 543.63 & 27.66 & 25.26 & 25.64 & 26.62 \\
 \  MB575 & 575.64 & 34.72 & 24.91 & 25.28 & 26.26 \\
 \  MB611 & 611.40 & 36.78 & 24.66 & 25.04 & 26.02 \\
 \  MB648 & 648.62 & 36.04 & 24.71 & 25.09 & 26.07 \\
 \  MB685 & 685.92 & 38.53 & 24.64 & 25.02 & 26.00 \\
 \  MB725 & 725.58 & 40.76 & 24.70 & 25.08 & 26.06 \\
 \  MB767 & 767.53 & 43.13 & 24.17 & 24.55 & 25.52 \\
 \  MB811 & 811.92 & 45.61 & 24.16 & 24.54 & 25.58 \\
 \  MB858 & 858.67 & 47.85 & 24.02 & 24.40 & 25.43 \\
 \  MB908 & 908.24 & 49.66 & 23.74 & 24.12 & 25.15 \\
 \  MB959 & 959.54 & 52.91 & 23.38 & 23.75 & 24.79 \\
\bottomrule
\end{tabular*}
\tablefoot{The table includes limiting depths for different exposure times.}
\label{tab:MBfilters}
\end{table*}
Analogously to the steps followed across the analyses presented in this paper, we calculate the bias in the SED metric for a set of 60\,000 simulated galaxies, and we study its trend as a function of redshift. The results, reported in \cref{fig:08}, show that an even sampling of the \IE filter, not exceeding the rate of $\sim 30$\,nm, leads both TFC and GP to a more accurate and robust SED reconstruction. Similar behaviour is observed in the application of interpolation methods, as displayed in \cref{fig:11}. In this case, the dispersion of the bias estimates per redshift bin is reduced to an average of 0.005.

\subsection{Future perspectives}
\subsubsection{Machine-learning-based template adaptation}
An alternative approach worth exploring is to model the residuals between the observed fluxes and the TF predictions using GP. In this method, GP regression is set to capture finer deviations from typical galaxy features not well represented in the templates, but still preserving the overall SED shape. This method may lead to a better reconstruction of sharp spectral features, although it remains subject to the same limitations with respect to broad-band smoothing. We leave a detailed investigation of this adaptive correction strategy to future work.

\subsubsection{Other sources of bias}
In relation to the analysis presented in this work, we comment on three additional sources of uncertainty. 
First, sample variance may arise when reference samples are built from limited sky areas such as COSMOS, which may lack a complete representation of the overall galaxy population. On the other hand, as emphasised by EH18, using the same photometric data for both for photo-$z$ and for the SED reconstruction can lead to covariance.

The \Euclid PHZ PF plans to mitigate the former by extending the ancillary data to the Euclid Auxiliary Fields (EAFs), while the latter requires a dedicated treatment of the joint inference of redshift and SEDs, which we leave for future work.

An additional source of bias is expected to arise from matching the EWS galaxies with the reference sample through NNPZ. Studying this effect is beyond the scope of the present analysis and will be assessed in a dedicated future study. The potential degradation introduced via the NNPZ step further highlights the importance of the new medium-band survey presented in \cref{sec:MBnew}, as it can reduce the bias already at the SED reconstruction stage.



\section{Conclusions}

The billions of galaxies that the \Euclid mission will observe are set to be part of an unprecedented study of the large-scale structure of the Universe. Through the \IE broad-band filter, \Euclid will measure their shapes with high accuracy. However, these measures strongly depend on an accurate PSF model, which is linked to the galaxy SEDs. These are reconstructed using supporting photometric datasets that provide point fluxes from available medium and broad-band filters. However, the uneven sampling of the wavelength range covered by the \IE filter can impact the reconstruction process and inaccuracies may lead to bias in the PSF model and thus in galaxy shape measurements. Studying this inaccuracy is key to minimise the bias introduced in the \Euclid core science analyses and ensure robust constraints on cosmological parameters.

Our research focuses on the analysis of the bias in the reconstruction of galaxy SEDs for the \Euclid mission. We introduce a novel SED metric, $\mathcal{M}_{\lambda}$, interpreted as the effective central wavelength of the \IE pass-band for a given SED, and defined through the same formalism that describes the PSF quadrupole moments, quantifying the PSF size and ellipticity metrics. This ensures a direct translation of the bias in the SED metric into biases in the PSF model. According to C13, the requirement on the fractional bias in the \Euclid PSF size metric due to wavelength dependency should remain below $3.5\times 10^{-4}$, in order to avoid the risk of systematics on the PSF affecting cosmic shear analyses.

In this work, we explore different SED reconstruction methods, including TFC and GP, and evaluate if their performance meets the requirements imposed on the \Euclid PSF model.
Our results reveal that GP can provide accurate SED reconstruction in specific redshift intervals, while TFC exhibits robustness in other ranges, where the typical features of galaxy SEDs fall within the \IE window. Combining the strengths of both methods, we propose a combined approach that meets the \Euclid requirements across almost all redshift bins. This methodological framework is adopted in the PHZ PF to reconstruct galaxy SEDs for the \Euclid survey

Furthermore, we exploit the possibility of using a newly developed filter design to improve SED reconstruction in future analyses. This new set of HSC-MB filters consists of 16 evenly spaced imaging filters which provides a quasi-spectroscopic uniform sampling of the \IE wavelength range. SED reconstruction applied on simulations based on this design demonstrates promising results, achieving an accuracy level that fulfils the \Euclid requirements across all redshift bins.

In summary, our findings display the potential of combining physics- and data-driven approaches, offering a comprehensive strategy for achieving galaxy SED reconstruction that is close to fully meeting the requirements for the \Euclid science mission. Notably, our results highlight that a well-designed set of evenly spaced HSC-MB filters can lead to overcome the current limitations in the SED sampling, and improve the accuracy and robustness of the SED reconstruction across all redshifts. This new infrastructure would benefit the \Euclid core science, paving the way for future precision weak-lensing analyses.

\begin{acknowledgements}
\AckEC 
\end{acknowledgements}

%
%

\bibliography{Euclid}

%
%

\begin{appendix}
\clearpage 
\newcommand{\refer}[1]{,\\ \cite{#1}; [{\tt #1}]}
\newcommand{\itemm}{

\medskip\noindent}

\section{Regression with Gaussian processes}
\label{sec:AppGP}
GP is a machine-learning algorithm used for both classification and regression tasks. As mentioned in \cref{sec:S4}, this method does not infer a distribution over the parameters of a specific function. Instead, it is used to infer a distribution over functions. More precisely, the process works by defining a prior distribution over functions, and updating the distribution based on observed data to obtain a posterior distribution. This section describes the most salient steps covered by this process, illustrating examples recreated from the astronomical dataset used in this work. For a full treatment of the topic we refer the reader to \citet{10.7551/mitpress/3206.001.0001}.

\begin{figure}
\includegraphics[width=\columnwidth]{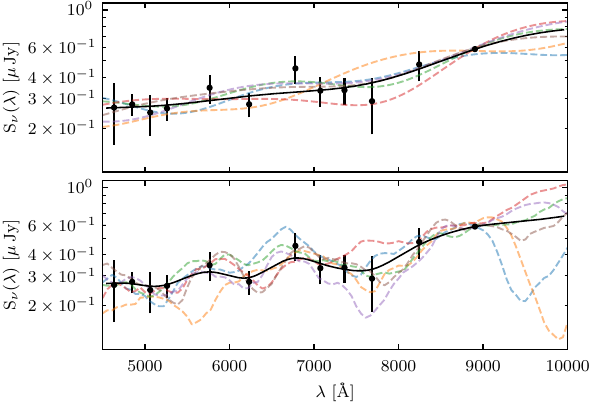}
\caption{Example of SED reconstruction through GP regression for a randomly selected simulated galaxy, using the RBF kernel (top panel) and the Matérn kernel (bottom panel). The key components of the regression are illustrated, with six functions sampled from the posterior distribution (dashed coloured lines), and the mean predictive function (solid black line).}
\label{fig:kernels}
\end{figure}

Each point in the dataset has an input value $\mathbf{x} = \{x_1, x_2, ..., x_n \}$ corresponding to an observation $\mathbf{y} = \{y_1, y_2, ..., y_n \}$. The goal of GP regression is to calculate a function $f_*$ assigning an observation $y_{*} = f_*(x_*)$ to each new input $x_{*}$. GP associates any point $\mathbf{x}$ to a random variable, $f(\mathbf{x})$, whose $N$-dimensional joint distribution, $p(f(\mathbf{x_1}), f(\mathbf{x_2}), ..., f(\mathbf{x_N}))$, is Gaussian and defined as
\begin{equation}
    p(\mathbf{f}|\mathbf{X}) = \mathcal{N}(\mathbf{f}|\mathbf{\mu}, \mathbf{K}) \, ,
\label{eq:A1}
\end{equation}
where $\mathbf{f} = f(\mathbf{x_1}), f(\mathbf{x_2}), ..., f(\mathbf{x_n})$, $\mathbf{\mu} = (m(\mathbf{x_1}), m(\mathbf{x_2}), ..., m(\mathbf{x_n}))$ and $K_{ij} = \kappa(x_i, x_j)$. More precisely, $\mathbf{\mu}$ is the sample mean function and $\kappa$ is a positive definite covariance function or \textit{kernel}. The covariance function quantifies the similarities between the points, and propagates the information into their respective observations. This means that if $x_i$ and $x_j$ are similar, the values of $f(x_i)$ and $f(x_j)$ are also expected to be similar. Therefore the kernel is responsible for shaping the function in terms of linearity and smoothness. The predictive distribution is given by
\begin{equation}
p(\mathbf{f_*}|\mathbf{X_*},\mathbf{X},\mathbf{f}) = \mathcal{N}(\mathbf{f_*}|\mathbf{\mu_*}, \mathbf{\Sigma_*}) \, ,
\end{equation}
where $\mathbf{\Sigma_*}$ summarises the covariance between the original input data points and the new inputs. For the latter, predictions can be made through the posterior distribution by computing the mean and variance of the distribution itself at those points. The mean provides the predicted function value, while the variance gives an estimate of the uncertainty in the prediction. In \cref{fig:kernels} we can visualise all the concepts discussed so far. The SED of a simulated galaxy is reconstructed with GP from the set of Subaru SC medium-bands described in \cref{sec:4.2}. First, GP builds a distribution of prior functions, $p(\mathbf{f}|\mathbf{X})$. Then the prior distribution is updated according to the function values and the correlation and the noise level of the flux points. Six functions, randomly sampled from the posterior distribution, and the mean function of the posterior distribution, $f_*$, are also displayed. The latter can be evaluated at any new wavelength, $\lambda_*$, falling within the $\lambda$ regression range, to infer the value of the SED at that wavelength, $f_*(\lambda_*)$, with an error equal to the variance of the posterior. The GP regression is carried using a different choice of kernels, RBF in the top panel, and the Matérn kernel in the bottom panel\footnote{For further information we refer the reader to \url{https://scikit-learn.org/stable/modules/gaussian_process.html}.}. The RBF kernel is given by 
\begin{equation}
    \kappa(x_i, x_j) = \exp\left(-\frac{d^2(x_i, x_j)}{2\ell^2}\right) \, ,
\label{eq:RBF}
\end{equation}
where $d^2(x_i, x_j)$ is the Euclidean distance. This kernel is infinitely differentiable. This property  allows the GP to have mean square derivatives of all orders. Therefore these functions are extremely smooth.
The Matérn kernel is defined as
\begin{equation}
    \kappa(x_i, x_j) = \frac{1}{\Gamma(\nu)2^{\nu -1}} 
    \left(\frac{\sqrt{2\nu}}{\ell} d(x_i, x_j) \right)^{\nu} 
    K_{\nu} \left(\frac{\sqrt{2\nu}}{\ell} d(x_i, x_j) \right) \, .
\label{eq:Matern}
\end{equation}
The $\nu$ parameter models the smoothness of the predictive function. It is modelled by a length-scale parameter, $\ell$. This allows the kernel to recover more realistically the minute structure of the function and the similarities between the data points. As we observed in our tests and illustrated in \cref{fig:kernels}, the Matérn kernel is more suited than RBF for the task of galaxy SED reconstruction. 

\end{appendix}

\end{document}